\documentclass[aps,prd,onecolumn,showpacs,eqsecnum,nofootinbib]{revtex4}     

\usepackage{epsfig}
\usepackage{graphicx}
\usepackage{dcolumn}
\usepackage{amsmath}
\usepackage{enumerate}

\def\beq{\begin{equation}}
\def\eeq{\end{equation}}

\def\gtap{\mathrel{ \rlap{\raise 0.511ex \hbox{$>$}}{\lower 0.511ex
   \hbox{$\sim$}}}} 
\def\ltap{\mathrel{ \rlap{\raise 0.511ex
    \hbox{$<$}}{\lower 0.511ex \hbox{$\sim$}}}}
\newcommand{\deltaunodue}{\mbox{$\Delta m_{21}^2 $}}
\newcommand{\deltaunotre}{\mbox{$\Delta m_{31}^2 $}}

\newcommand{\tetaot}{\mbox{$\theta_{13}$}}
\newcommand{\probdiff}{\mbox{${\cal{D}}$}}
\newcommand{\nova}{\mbox{NO$\nu$A} }
\newcommand{\pbarnu}{\nu \hspace{-2.5mm}         
\raisebox{1.3ex}{{\tiny(}}\raisebox{1ex}{--}\raisebox{1.3ex}{\tiny)}}
\newcommand{\pbarP}{P \hspace{-3mm}
\raisebox{1.9ex}{{\tiny(}}\raisebox{1.4ex}{--}\raisebox{1.9ex}{\tiny)}
   }

\begin{document}

\vskip-6pt \hfill {CERN-PH-TH/2005-050} \\
\vskip-6pt \hfill {FERMILAB-PUB-05-050-T} \\

\title{
\vskip-12pt~\\
Super-NO$\nu$A: a long-baseline neutrino experiment with two off-axis
detectors}

\author{
\mbox{Olga Mena Requejo$^{1}$},
\mbox{Sergio Palomares-Ruiz$^{2}$} and
\mbox{Silvia Pascoli$^{3}$}}

\affiliation{
\mbox{$^1$ Theoretical Physics Department, Fermi National Accelerator
Laboratory, Batavia, IL 60510-0500, USA}
\mbox{$^2$ Department of Physics and Astronomy, Vanderbilt University,
  Nashville, TN 37235, USA}
\mbox{$^3$ Physics Department, Theory Division, CERN, CH-1211 Geneva
23, Switzerland}  
\\
{\tt omena@fnal.gov},
{\tt sergio.palomares-ruiz@vanderbilt.edu},
{\tt Silvia.Pascoli@cern.ch}
}

\begin{abstract}

Establishing the neutrino mass hierarchy is one of the fundamental
questions that will have to be addressed in the next future. Its
determination could be obtained with long-baseline experiments but
typically suffers from degeneracies with other neutrino parameters.
We consider here the \nova experiment configuration and propose to
place a second off-axis detector, with a shorter baseline, such that,
by exploiting matter effects, the type of neutrino mass hierarchy could
be determined with only the neutrino run. We show that the
determination of this parameter is free of degeneracies, provided the
ratio $L/E$, where $L$ the baseline and $E$ is the neutrino energy, is
the same for both detectors.
\end{abstract}


\maketitle

\section{Introduction}

During the last several years the progress in the studies of neutrino
oscillations has been remarkable. The experiments with
solar~\cite{sol,SKsolar,SNO1,SNO2,SNO3,SNOsalt},
atmospheric~\cite{SKatm}, reactor~\cite{KamLAND} and recently also
long-baseline accelerator~\cite{K2K} neutrinos have provided
compelling evidence for the existence of neutrino oscillations driven
by non-zero neutrino masses and neutrino mixing. We know that there
are two large ($\theta_{12}$ and $\theta_{23}$) and one small
($\theta_{13}$) angles, and at least two mass square
differences~\footnote{We restrict ourselves to a three-family neutrino
scenario analysis. The unconfirmed LSND signal~\cite{LSND} cannot be
explained in terms of neutrino oscillations within this scenario, but
might require additional light sterile neutrinos or more exotic
explanations (see e.g. Ref.~\cite{LSNDexpl}). The ongoing  MiniBooNE
experiment~\cite{miniboone} is expected to explore all of the LSND
oscillation parameter space~\cite{LSND}.}, $\Delta m_{ji}^{2} \equiv
m_j^2 -m_i^2$, with $m_{j,i}$ the neutrino masses, one associated
to atmospheric neutrino oscillations ($\deltaunotre$) and one to solar
ones ($\deltaunodue$). The angles $\theta_{12}$ and $\theta_{23}$
represent the neutrino mixing angles responsible for the solar and the
dominant atmospheric neutrino oscillations, while $\theta_{13}$ is the
angle limited by the data from the CHOOZ and Palo Verde
experiments~\cite{CHOOZ,PaloV}.   

Stronger evidences of neutrino oscillations are provided by the new
Super-Kamiokande data on the $L/E$ dependence of multi-GeV $\mu$-like
atmospheric neutrino events~\cite{SKdip04}, $L$ being the distance
traveled by neutrinos and $E$ the neutrino energy, and by the new
more precise spectrum data of the KamLAND~\cite{KL766} and K2K
experiments~\cite{K2K}. For the first time these data exhibit
directly, not only a deficit with respect to the expected signal, but
also the effects of the oscillatory dependence on $L/E$ of the
probabilities of neutrino oscillations in vacuum~\cite{BP69}. We begin
to actually ``see'' the oscillatory behavior of neutrino propagation. 

The Super-Kamiokande and K2K data are best described in terms of
dominant $\nu_{\mu} \rightarrow \nu_{\tau}$ ($\bar{\nu}_{\mu} 
\rightarrow \bar{\nu}_{\tau}$) vacuum oscillations. The best-fit
values explaining the Super-Kamiokande data~\cite{SKatm} are
$|\deltaunotre| = 2.1\times 10^{-3}~{\rm eV^2},~\sin^22\theta_{23} =
1.0~$, whereas those for the K2K data~\cite{K2K} are $|\deltaunotre|
= 2.8\times 10^{-3}~{\rm eV^2},~\sin^22\theta_{23} = 1.0~$. The
90\% C.L. allowed ranges of the atmospheric neutrino oscillation
parameters obtained by the Super-Kamiokande experiment
read~\cite{SKatm}:  
\beq 
\label{eq:range}
|\deltaunotre| =(1.5 - 3.4)\times10^{-3}{\rm eV^2},~~~~
\sin^22\theta_{23}\geq 0.92.
\eeq
The sign of $\deltaunotre$ and of $\cos2\theta_{23}$, when
$\sin^22\theta_{23} \neq 1.0$, cannot be determined with the existing
data. For the mass square difference, the two possibilities,
$\deltaunotre > 0$ or $\deltaunotre < 0$, correspond to two different
types of neutrino mass ordering: normal hierarchy (NH), $m_1 <
m_2 < m_3$ ($\deltaunotre > 0$), and inverted hierarchy (IH),
$m_3 < m_1 < m_2$ ($\deltaunotre < 0$). The fact that the sign of
$\cos2\theta_{23}$ is not determined when $\sin^22\theta_{23} \neq
1.0$ implies that the octant where $\theta_{23}$ lies is not known.

In addition, the combined 2-neutrino oscillation analysis of the solar
neutrino data, including the results from the complete salt phase of
the Sudbury Neutrino Observatory (SNO) experiment~\cite{SNOsalt}, 
and the recent KamLAND 766.3 ton-year spectrum data~\cite{KL766} shows
that the solar neutrino oscillation parameters 
lie in the low-LMA (Large Mixing Angle) region, 
with the best fit value at
\cite{SNOsalt}
\beq
\deltaunodue =8.0 \times 10^{-5}~{\rm eV^2},~~~
\sin^2 \theta_{12} =0.31.
\eeq
In a 3-neutrino oscillation framework, a combined analysis of the
solar, atmospheric, reactor and long-baseline neutrino data 
gives~\cite{GG04} (see also Ref.~\cite{MSTV04}):   
\beq
\sin^2\theta_{13} < 0.041,~~~~3\sigma~{\rm C.L.} 
\label{eq:chooz}
\eeq
As we know that neutrinos do oscillate, there are very important
questions that will have to be addressed in future experiments. 
Besides the more accurate determination of the leading neutrino
oscillation parameters that will be achieved by the
MINOS~\cite{MINOS}, OPERA~\cite{OPERA} and ICARUS~\cite{ICARUS}
experiments and future atmospheric and solar neutrino
detectors~\cite{SAWG}, one of the most important tasks in the next
future will be the determination of the (1,3) sector of the lepton
mixing matrix, the Pontecorvo--Maki--Nakagawa--Sakata (PMNS) neutrino
matrix~\cite{BPont57}. A complete determination of this sector entails
the measurement of a non-zero $\theta_{13}$, which will open the door
to the experimental measurement of the CP-- (or T--) violating phase,
$\delta$, and to establishing the type of neutrino mass spectrum. 
This mixing angle controls the $\nu_\mu \rightarrow \nu_e$ and
$\bar{\nu}_\mu \rightarrow \bar{\nu}_e$ conversions in long-baseline
experiments and in the widely discussed very long-baseline neutrino
oscillation experiments at neutrino factories~\cite{nufact,CDGGCHMR00}.
More recently, $\beta$-beams experiments, exploiting neutrinos from
boosted-ion decays~\cite{zucchelli,mauro,p2}, with an improved
experimental setup have been shown to achieve sensitivities to
leptonic CP--violation and to the sign of the atmospheric mass
difference competitive with those at neutrino factories~\cite{p1}.  
The mixing angle $\theta_{13}$ also controls the Earth matter effects
in multi-GeV
atmospheric~\cite{atmmatter1,mantle,core,atmmatter2,atmmatter3} and in
supernova~\cite{SN} neutrino oscillations. Finally, the magnitude of
the T-violating and CP--violating terms in neutrino oscillation
probabilities is directly proportional to $\sin\theta_{13}$~\cite{CPT}.
Therefore the determination of the magnitude of $\theta_{13}$ is
crucial for the future searches for matter effects and CP--violation in
the lepton sector at neutrino oscillation experiments.

The measurement of, or a stronger limit on, the mixing angle
$\theta_{13}$ in the near  future is going to be achieved by 
reactor~\cite{reactor} and
long-baseline~\cite{MINOS,OPERA,ICARUS,T2K,NOvA,newNOvA} neutrino  
experiments. Neutrino reactor experiments, being disappearance
experiments, are only sensitive to the value of $\theta_{13}$. 
Long-baseline neutrino experiments, in addition to having a better
sensitivity to $\theta_{13}$, are the only way (in the near future) to
search for CP violation, while being able to determine the type of
hierarchy at the same time\footnote{The same capabilities with better
sensitivity could be achieved by neutrino
factories~\cite{nufact,CDGGCHMR00}. Information on the type of
neutrino mass hierarchy might be obtained in future atmospheric
neutrino experiments~\cite{mantle,core,atmmatter2,atmmatter3}. If
neutrino are Majorana particles, next generation neutrinoless double
$\beta-$decay experiments could establish the type of neutrino mass
spectrum~\cite{PPRmass} (see also
Refs.~\cite{Bilmass,BPP,PPW,betabetamass}) and, possibly, might 
provide some information on the presence of CP--violation in the lepton
sector due to Majorana CP--violating phases~\cite{PPRCP} (see also
Refs.~\cite{BilCPV,BPP,PPW,betabetaothers}).}. However, long-baseline
neutrino oscillation experiments suffer from degeneracies in the
neutrino parameter space~\cite{FL96,BCGGCHM01,MN01,BMWdeg,deg}.
In general, the proposed experiments have a single detector with the
beam running in two different modes, neutrinos and
antineutrinos. With only one neutrino and one antineutrino run, the
degeneracies can lead to different CP--violating and CP--conserving
sets of parameters explaining the data at the same confidence level. 
In Ref.~\cite{CDGGCHMR00} it was pointed out that some of the
degeneracies could be eliminated with sufficient energy or baseline
spectral information. In practice, however, the spectral information
has been shown to be not strong enough to resolve degeneracies with a
single detector, once that statistical errors and realistic
efficiencies and backgrounds are taken into account. In order to
resolve the parameter degeneracy, another
detector~\cite{BNL,MN97,BCGGCHM01,silver,BMW02off} or the combination 
with another
experiment~\cite{BMW02,HLW02,MNP03,otherexp,mp2,HMS05,unveiling}
would, thus, be necessary. Recently, new approaches for determining
the type of hierarchy have been proposed~\cite{hieratm}; they
typically exploit other neutrino oscillations channels, such as muon
neutrino disappearance, and require very precise neutrino oscillation
measurements. 

Contrary to the na\"{\i}ve expectation, it has been shown numerically
in Ref.~\cite{HLW02} and analytically in Ref.~\cite{MNP03} that the use
of only a neutrino beam could help in resolving the type of hierarchy
when two different long-baseline experiments are combined under 
certain conditions.

Differently from this approach, we present here a scenario with only
one experiment, which runs in the neutrino mode and uses two detectors
at different distances and different off-axis angles. It is well known
that off-axis neutrino beams have a very narrow neutrino spectra, and
that the peak energy can be tuned by just moving the detector with
respect to the main beam axis. We notice that an off-axis beam can be
obtained by either displacing the detector a few km away from the
location of an on-axis surface detector, or by placing it on the
vertical of the beam-line but at a much shorter distance. In such a
way, a single beam could do the job of two beams with different
energies.

We study the use of two off-axis detectors in combination with the
NuMI beam so that the type of mass spectrum could be determined free
of other degeneracies, if $\theta_{13}$ is not very small. We will
consider, for one of them, the location which is the most likely for
the \nova configuration\footnote{This values correspond to the already
old \nova proposal~\cite{NOvA}. For the recent revised proposal see
Ref.~\cite{newNOvA}.} ($L = 810$ km and $E = 2.3$ GeV) and show
different possibilities for the baseline of the other detector in
order to make the measurement of $sign(\deltaunotre)$ feasible with
only the neutrino beam. We name this improved experimental setup 
Super-$\nova$~\cite{cooper}. Following the line of thought of
Ref.~\cite{MNP03}, we will show that a configuration with the same
vacuum oscillation phase, i.~e. same $L/E$ for both detectors, is
specially sensitive to matter effects. For such an experimental setup
the sensitivity to $sign(\deltaunotre)$ would be enhanced as the
difference in baseline lengths grows. This configuration also has the
advantage of requiring only one experiment and of reducing the error
due to systematic uncertainties from the beam. In addition, we will
show that such a measurement is free of degeneracies, which provides
the full power of this method. We start by presenting the general
formalism in Sec.~\ref{formalism}. In Sec.~\ref{setup} we describe the
experimental setup. We explain in Sec.~\ref{matter} the method to
extract the type of neutrino mass spectrum free of degeneracies by
using this special configuration and we show how the sensitivity
changes for different values of $|\deltaunotre|$. Finally, in
Sec.~\ref{conclusions}, we make our final remarks. In
Appendix~\ref{appendix} we present the computed charged-current (CC)
neutrino event rates for a particular choice of parameters.

\section{Formalism}
\label{formalism}

We consider the probability of $\pbarnu_\mu \rightarrow \pbarnu_e$
oscillation, $P\left(\pbarnu_\mu \rightarrow \pbarnu_e; L\right) \equiv
\pbarP(L)$, in the context of three-neutrino mixing. For neutrino
energies $E \gtap$ 1 GeV, $\theta_{13}$ within the present
bounds~\cite{GG04,MSTV04}, and baselines $L \ltap 1000$
km~\cite{BMWdeg}~\footnote{For $E \gtap 0.6$ GeV we have checked that
the analytical expansion is accurate for $L < 500$ km within the
present bounds of $\theta_{13}$ and $\deltaunodue/\deltaunotre$.}, the
oscillation probability $\pbarP(L)$ can be safely approximated
by expanding in the small parameters $\theta_{13}$,
$\Delta_{12}/\Delta_{13}$, $\Delta_{12}/A$ and $\Delta_{12} L$ , where
$\Delta_{12} \equiv \deltaunodue/(2 E)$ and $\Delta_{13} \equiv
\deltaunotre/(2E)$~\cite{CDGGCHMR00} (see also Ref.~\cite{3prob}):     
\beq
\begin{array}{ll}
\pbarP(L) \simeq & 
\sin^2 \theta_{23} \, \sin^2 {2 \theta_{13} } \left(
\frac{\Delta_{13}}{A \mp \Delta_{13}} \right)^2     
\sin^2 \left( \frac{(A \mp \Delta_{13}) L}{2} \right) \\
& + \cos \theta_{13} \sin {2 \theta_{13} } \sin {2 \theta_{23}} 
\sin {2 \theta_{12}} \ \frac{\Delta_{12}}{A} \frac{\Delta_{13}}{A \mp
  \Delta_{13}} \ 
\sin \left(\frac{A L}{2} \right) \sin \left( \frac{(A \mp
  \Delta_{13}) L}{2} \right) 
 \cos \left(\frac{\Delta_{13} L}{2} \mp \delta \right) \\
 & + \cos^2 \theta_{23} \sin^2 {2 \theta_{12}} \left(
 \frac{\Delta_{12}}{A} \right)^2 \sin^2 \left( \frac{A L}{2} \right), 
\end{array}
\label{eq:probappr}
\eeq
where $L$ is the baseline. We use the constant density approximation
for the index of refraction in matter $A$, defined as $A \equiv
\sqrt{2} G_{F} \bar{n}_e(L)$, with $\bar{n}_e(L)$ the average electron
number density, defined by $\bar{n}_e(L) = 1/L \int_{0}^{L} n_e(L')
dL'$. Here $n_e(L)$ is the electron number density along the baseline. 

As is well known~\cite{MN01}, the CP trajectory in bi--probability
plots of neutrino and antineutrino conversion at the same baseline,
 is elliptic under the assumption of mass hierarchy and of
adiabaticity. The ellipses obtained for each of the two hierarchies,
and for different values of $\delta$, $\theta_{13}$ and of
$\theta_{23}$, intersect in points that correspond to 2--, 4--, and
8--fold degeneracies~\cite{FL96,BCGGCHM01,MN01,BMWdeg,deg}. It follows
that even a precise determination of a point in the $P$--$\bar{P}$
plane can result in different sets of CP--conserving and CP--violating
parameters $(\delta, \, \theta_{13}, \, \theta_{23}, \,
sign(\deltaunotre) )$, all of which reproduce the observations. The
allowed regions in the $P$--$\bar{P}$ plane obtained by varying the
values of $\theta_{13}$ and $\delta$ within their ranges describe wide
``pencils''. The ``pencils'' for the cases of normal and inverted
hierarchy have different slopes and overlap for a large fraction. This
indicates that, generically, a measurement of the probability of
conversion for neutrinos and antineutrinos cannot uniquely determine
the type of hierarchy in a single experiment. In order to resolve such
degeneracy, various strategies have been proposed: combined analysis
of the data from different super-beam experiments, or of the data from
super-beam facilities and neutrino factories (or
$\beta$-beams)~\cite{p1,BMW02,HLW02,MNP03,otherexp,mp2}, the use of
additional information from atmospheric neutrino data~\cite{HMS05},
and experimental setups with clusters of
detectors~\cite{BNL,MN97,BCGGCHM01,BMW02off}. 

It has been pointed out that considering the probabilities of neutrino
oscillations only, at two different baselines and energies, can resolve
the type of hierarchy~\cite{HLW02,MNP03}. In the case of
bi--probability plots of neutrino--neutrino conversions at different
baselines, the CP--trajectory is elliptic too. Again, the allowed
regions in these bi--probability plots form two ``pencils'', which grow
in width away from the origin, each of them associated with one type
of mass spectrum. The overlap of the two ``pencils'', which signals
the presence of a degeneracy of the type of hierarchy with other
parameters, is controlled by the slope and the width of the
``pencils''. From Eq.~(\ref{eq:probappr}) one can see that the ratio of
the slopes of the central axes of these two ``pencils'' in the $P_{\rm
F}$--$P_{\rm N}$ plane, where $P_{\rm F}$ ($P_{\rm N}$) is the
neutrino conversion probability at the far (near) detector, is given
by~\cite{MNP03} 
\beq
\frac{\alpha_+}{\alpha_-} = 
\frac{
\left(\frac{\Delta_{13,{\rm N}}}{A_{\rm N} - \Delta_{13,{\rm N}}}
\right)^2  
\sin^2 \left( \frac{(A_{\rm N} - \Delta_{13,{\rm N}}) L_{\rm N}}{2}
\right)  
\left(\frac{\Delta_{13,{\rm F}}}{A_{\rm F} + \Delta_{13,{\rm F}}}
\right)^2      
\sin^2 \left( \frac{(A_{\rm F} + \Delta_{13,{\rm F}}) L_{\rm F}}{2}
\right)} 
     {\left(\frac{\Delta_{13,{\rm F}}}{A_{\rm F} - \Delta_{13,{\rm
	     F}}} \right)^2  
\sin^2 \left( \frac{(A_{\rm F} - \Delta_{13,{\rm F}}) L_{\rm F}}{2}
\right)  
\left(\frac{\Delta_{13,{\rm N}}}{A_{\rm N} + \Delta_{13,{\rm N}}}
\right)^2      
\sin^2 \left( \frac{(A_{\rm N} + \Delta_{13,N}) L_{\rm N}}{2} \right)}
\label{ratio}
\eeq
where $\alpha_+$ and $\alpha_-$ are the slopes for normal and inverted
hierarchy, respectively; $\Delta_{13,{\rm F(N)}}$, $A_{\rm F(N)}$ and
$L_{\rm F(N)}$ are the values of $\Delta_{13}$, $A$ and $L$ for the
far (near) detector. Note that although we are using the constant
density approximation, $A_{\rm F}$ and $A_{\rm N}$ are different
because the average density depends on the baseline. Using the fact
that matter effects are small ($A\ll\Delta_{13}$), we can perform a
perturbative expansion, which up to first order gives this ratio of
slopes as 
\beq
\frac{\alpha_+}{\alpha_-} \simeq
1 + 2 \, A_{\rm N} L_{\rm N} \left( \frac{1}{(\Delta_{13,{\rm N}}
L_{\rm N} / 2)} - 
\frac{1}{\tan(\Delta_{13,{\rm N}} L_{\rm N} / 2)} \right) - 
2 \, A_{\rm F} L_{\rm F} \left( \frac{1}{(\Delta_{13,{\rm F}} L_{\rm
F} / 2)} - \frac{1}{\tan(\Delta_{13,{\rm F}} L_{\rm F} / 2)} \right).
\label{ratioapp}
\eeq
For the case of $L/E$ constant, noting that $1/x - 1/\tan x$ is a
monotonically increasing function, we conclude that the smaller the
chosen energy, the larger the ratio of slopes. This ratio increases
also for certain configurations with different
$L/E$~\cite{MNP03}. Another very important feature is the width, which
is very small for equal $L/E$, but grows rapidly when this is not the
case~\cite{MNP03}. Hence, even when the separation between the central
axes of the two regions is substantial, unless the ratio $L/E$ is kept
close to constant, the ellipses overlap, making the discrimination of
$sign(\deltaunotre)$ challenging. As was shown in Ref.~\cite{MNP03},
away from the $L/E$-constant case, the choice $L_{\rm F}/E_{\rm F} >
L_{\rm N}/E_{\rm N}$ is preferred. Otherwise, no matter how accurate
the experiment is, the discrimination of the type of neutrino mass
hierarchy free of degeneracies will not be possible. As a matter of
fact, this is precisely what will happen when combining \nova and T2K
experiments, for which $L_{\rm{NO}\nu\rm{A}}/E_{\rm{NO}\nu\rm{A}} =
352 \, \rm{km}/\rm{GeV} <  421 \, \rm{km}/\rm{GeV} =
$$L_{\rm{T2K}}/E_{\rm{T2K}}$\footnote{This corresponds to the
configuration with an off-axis angle of $2^o$ (OA$2^o$) for
T2K~\cite{T2K}.}. In this case, a joint analysis of these two
experiments does not present any interesting synergy effects and just
accounts for adding in statistics~\cite{HLW02}.

Thus, we will consider the case of $L/E$ constant and show how, by
adding another detector to the already proposed \nova experimental
setup, the measurement of the $sign(\deltaunotre)$ is possible free of
degeneracies from other parameters.

\section{Experimental setup}
\label{setup}

As was pointed out in the previous section, we consider here only one
experiment by using the same beam but having two detectors. In order
to maximize the sensitivity to the type of hierarchy, we propose a
configuration in which the neutrino conversion takes place predominantly
at the same $L/E$ at the two locations. In such a way, matter effects
can be factored out and the type of neutrino mass hierarchy can be
determined (if $\theta_{13}$ is large enough), free of degeneracies
from other parameters of the neutrino mixing matrix.

In order to have the same $L/E$ for both detectors we would need a
very well peaked spectrum at both sites. This can be achieved by
placing the detectors off the central axis of the beam. As
is well known, most of the neutrinos in a conventional neutrino 
beam are produced in two-body decays $\pi^{\pm} \to \mu^{\pm} +
\pbarnu_\mu$. The energy and flux of these neutrinos is determined by
the decay angle $\theta$~\cite{adamoff}: 
\beq  
E = \frac{0.43 \, E_\pi}{1 + \gamma^2 \, \theta^2},
\label{eq:eoff}
\eeq 
\beq 
\Phi = \left(\frac{2 \, \gamma}{1 + \gamma^2 \, \theta^2}\right)^2
\frac{1}{4 \pi L^2}~, 
\label{eq:fluxes}
\eeq  
where $\gamma$ is the Lorentz factor of the pion, $\theta$ is the
angle between the pion and the neutrino directions, and $L$ is the
distance between the decay point and the detector. A neutrino beam
with narrow energy spectrum can be produced by placing the detector
off-axis, i.~e., at some angle with respect to the forward direction
$\theta_{\rm beam}=0$. By using off-axis beams, one manages a
kinematic suppression of the high energy neutrino components,
whereas the low energy flux is kept approximately the same as that of
the on-axis beams. The neutrino spectrum is very narrow in energy and
peaked at lower energies with respect to the on-axis one. The
suppression of the high-energy tail of the spectrum greatly reduces
the backgrounds due to neutral-current interactions and $\tau$
production. Since the neutrino flux is nearly monochromatic, the
off-axis technique allows a discrimination between the peaked $\nu_e$
oscillation signal and the intrinsic $\nu_e$ background which has a
broad energy spectrum. An efficient reduction of the intrinsic
background can therefore be achieved~\cite{adamoff}. 

The off-axis angle and baseline for each detector must be chosen in
such a way that we have the same $L/E$ at both sites. From
Eqs.~(\ref{eq:eoff}) and~(\ref{eq:fluxes}) we see that the flux scales
as $\Phi \sim (E/L)^2$; so for this special configuration the flux at
the near detector is of the same order as that at the far detector (see
Appendix~\ref{appendix}). For the far detector we will use the
configuration proposed for the \nova experiment and we will suggest to
place another detector of the same characteristics at a closer
distance from the source.    

From geometrical considerations, and using the fact that the Earth is
curved, a detector located on the Earth surface, on the vertical line
of the central axis of the beam, is off-axis by a small angle,
$\theta_{\rm min}$. This is the minimum off-axis angle at a given
distance for a given beam configuration, which for $L_{\rm 
F,N,{on-axis}} \ll R$, reads 
\beq
(\theta_{\rm min})_{\rm F,N} \simeq \frac{L_{\rm on-axis} - L_{\rm
    F,N}}{2 R}, 
\label{eq:offmin}
\eeq
where $L_{\rm on-axis} = 735$ km is the baseline for the on-axis
detector (MINOS), $R$ is the Earth radius, and we have neglected terms
of order $(L_{\rm F,N,{on-axis}}/R)^3$. A different (larger) off-axis
angle at the same distance can be achieved by placing the detector
slightly outside the vertical of the beam. We present here the possible
locations for the near detector. 

It turns out that the peak energy in the $\nu_\mu$ CC neutrino event
spectrum is well fitted by the parametrization~\cite{privatemessier}   
\beq
E_{\rm peak} = \frac{1900}{(\theta + 16)^2}~\rm{GeV},
\label{eq:epeak}
\eeq
with $\theta$ the off-axis angle in units of mrad. Then one has to
solve  for the detector locations that have a constant ratio of $L/E 
\simeq 810 \, \rm{km}/2.3 \, \rm{GeV}  =  352 \, \rm{km}/\rm{GeV}$. 
From Eq.~(\ref{eq:epeak}), it is clear that we can write $\theta$ as a
function of the baseline $L$ for constant $L/E$, which reads
\beq
\theta(L) = 16 \left( 51 \left(\frac{\rm{km}}{L}\right)^{1/2}
\left(\frac{L/E}{352 \, \rm{km}/\rm{GeV}}\right)^{1/2} - 1
\right) \rm{mrad}.
\label{eq:offl}
\eeq
Once the possible values of the off-axis angle $\theta$ and the
associate peak energy $E_{\rm peak}$ are determined for a given
distance $L$, it is absolutely necessary to check whether or not the
geometry of the Earth and the NuMI beamline allow the different
configurations to be a reality~\cite{privatemessier}. From
Eqs.~(\ref{eq:offmin}) and~(\ref{eq:offl}), the condition for this to
be possible is $\theta (L) > \theta_{\rm min}$. In particular, there
are no sites between 300 and 400 km which give an $L/E$ ratio of 352
km/GeV. We have explored four possible detector locations: 200, 434,
500 and 600 km~\cite{privatemessier}. Here we only present the results
for the detector locations at 200 km and 434 km, since the distinction
of the neutrino mass hierarchy is easier for shorter baselines, as we
will show below. These configurations correspond to off-axis angles of
42.2 and 23.6 mrad, respectively. We have used the neutrino and
antineutrino fluxes available at different off-axis configurations  
at the \nova far site (810 km) and obtained the fluxes at the short
baselines by a simple rescaling of the fluxes at the far
distance\footnote{The short-baseline fluxes at 200 km can be easily
obtained from the fluxes at 735 km and 30 km off
axis~\cite{privateadam}. The fluxes at 434 km have been computed from  
the ones at 810 km and 18 km off-axis.}.       

In the old proposal~\cite{NOvA}, the $\nova$ far detector at 810 km is
a 50 kton tracking calorimeter, and the efficiency for its accepting a
$\nu_e$ event from $\nu_\mu \to \nu_e$ oscillations is approximately
$21\%$. We have explored here the possibility of using a 50 kton
liquid argon TPC detector~\cite{flare}, for which the efficiency to
identify $\nu_e$ CC interactions is 90\% (i.~e., basically perfect
efficiency) and that the background is dominated by the intrinsic
$\nu_e$ and $\bar{\nu}_e$ components of the beam. The high detection
efficiency of a liquid argon detector makes its statistics equivalent
to that of a conventional detector, with the beam power upgraded with
the proton driver. In addition, the physics potential of these
detectors is remarkable, as supernova neutrino detectors, for proton
decay searches\footnote{If protons decay primarily into kaons ($p
\rightarrow K^+ + \nu$), the detection efficiency in
water-\v{C}erenkov detectors is relatively low, for kaons at these 
energies are below threshold.} and for studies of neutrinoless double
beta decay (see Ref.~\cite{flare} and references therein).

We have assumed that the number of protons on target per year is $3.7
\times 10^{20}$~\cite{NOvA} ($18.5 \times 10^{20}$ pot/yr with the
Proton Driver) and five years of neutrino running. The revised \nova
proposal~\cite{newNOvA} suggests a  number of protons on target per
year which has been upgraded to $6.5 \times 10^{20}$ ($25 \times
10^{20}$ with a Proton Driver) and a 30 kton detector with $24\%$
efficiency.

\subsection{Oscillated statistics}

For a given value of the oscillation parameters, we have computed the
expected number of electron events, $N_{\ell}$ detected at the
possible locations $\ell = {\rm N, F}$ (near/far sites). The
observable that we exploit, $N_{\ell}$, reads  
\beq
N_{\ell,\pm} = \int^{E_{\rm max}}_{E_{\rm min}} \; 
\Phi_{\ell,\nu}(E_\nu,L) \; \sigma_{\nu}(E_\nu) \;
 P_{\nu}(E_\nu, L, \theta_{13}, \delta, \deltaunotre,
{\alpha}) \; dE_{\nu} 
\eeq
where the sign +($-$) applies for the normal (inverted) hierarchies and
${\alpha}$ is the set of remaining oscillation parameters:
$\theta_{23}, \, \theta_{12}, \, \deltaunodue$ and the matter parameter
$A$ (which depend on the baseline under consideration), which are
taken to be known; $\Phi_{\ell,\nu}$ denotes the neutrino flux and
$\sigma_{\nu}$ the cross sections.  The neutrino fluxes are thus
integrated over a narrow energy window, where $E_{\rm min}$ and
$E_{\rm max}$ refer to the lower and upper energy limits respectively.

For our analysis, unless otherwise stated, we will use a
representative $|\deltaunotre| = 2.4 \times 10^{-3} \ \rm{eV}^2$,
which lies within the best-fit values for the
Super-Kamiokande~\cite{SKatm} and K2K~\cite{K2K} experiments. However,
we will also show how the results change for different values of this
parameter. For the rest of the parameters, $\theta_{23}$,
$\deltaunodue$ and $\theta_{12}$, we will use the best fit values
quoted in the introduction. We show in Appendix~\ref{appendix} the
expected number of signal and background events at the far location
($L=810$ km) and at two of the possible near locations ($L=200$ km and
$L=434$ km) for both hierarchies and four central values of the CP
phase $\delta=0,\frac{\pi}{2}, \pi,\frac{3\pi}{2} $ and $\sin^2 2
\theta_{13}=0.058$.

\begin{figure}[t]
\begin{center}
\begin{tabular}{ll}
\epsfig{file=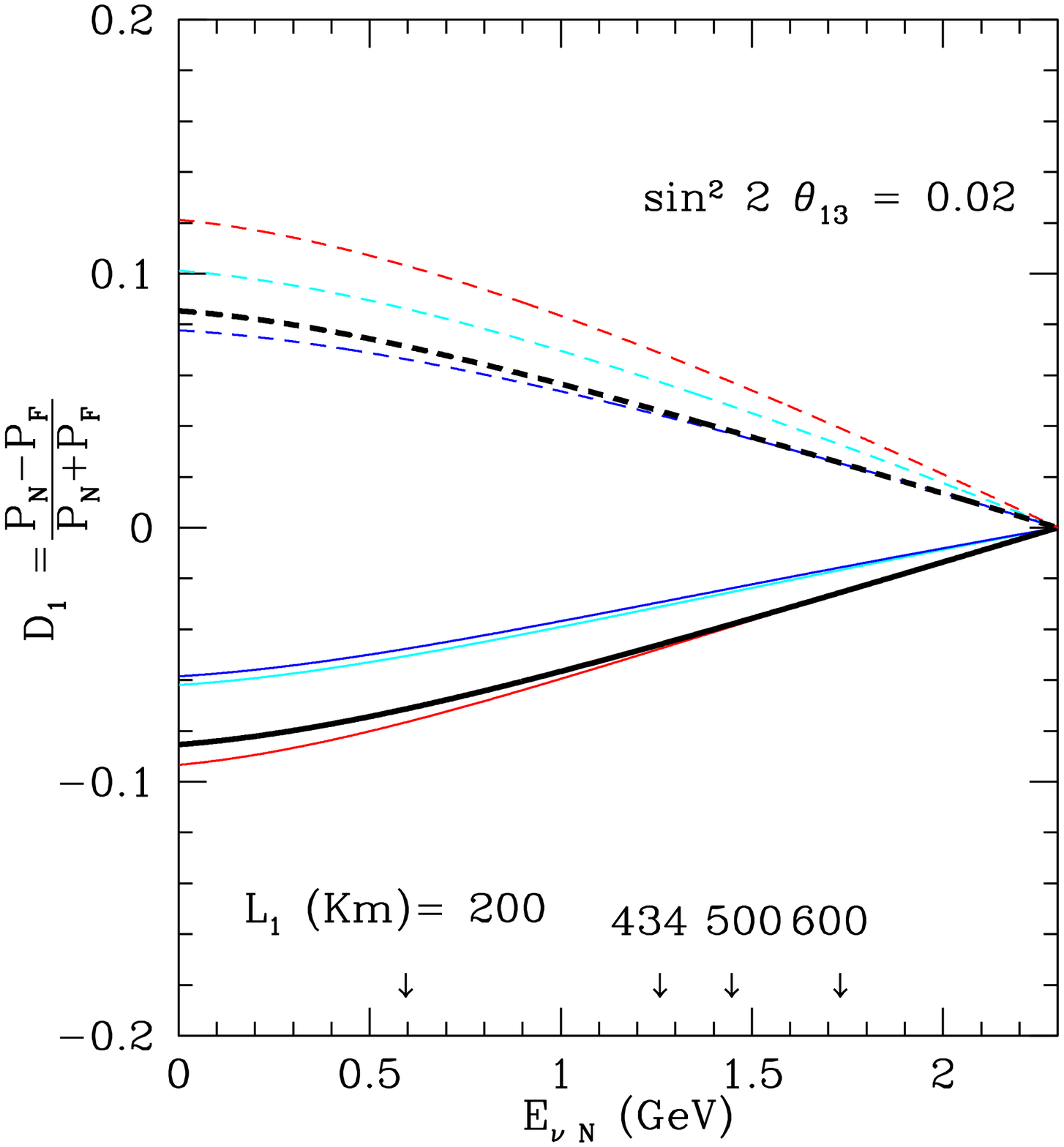, width=9.0cm} &
\hskip 0.cm
\epsfig{file=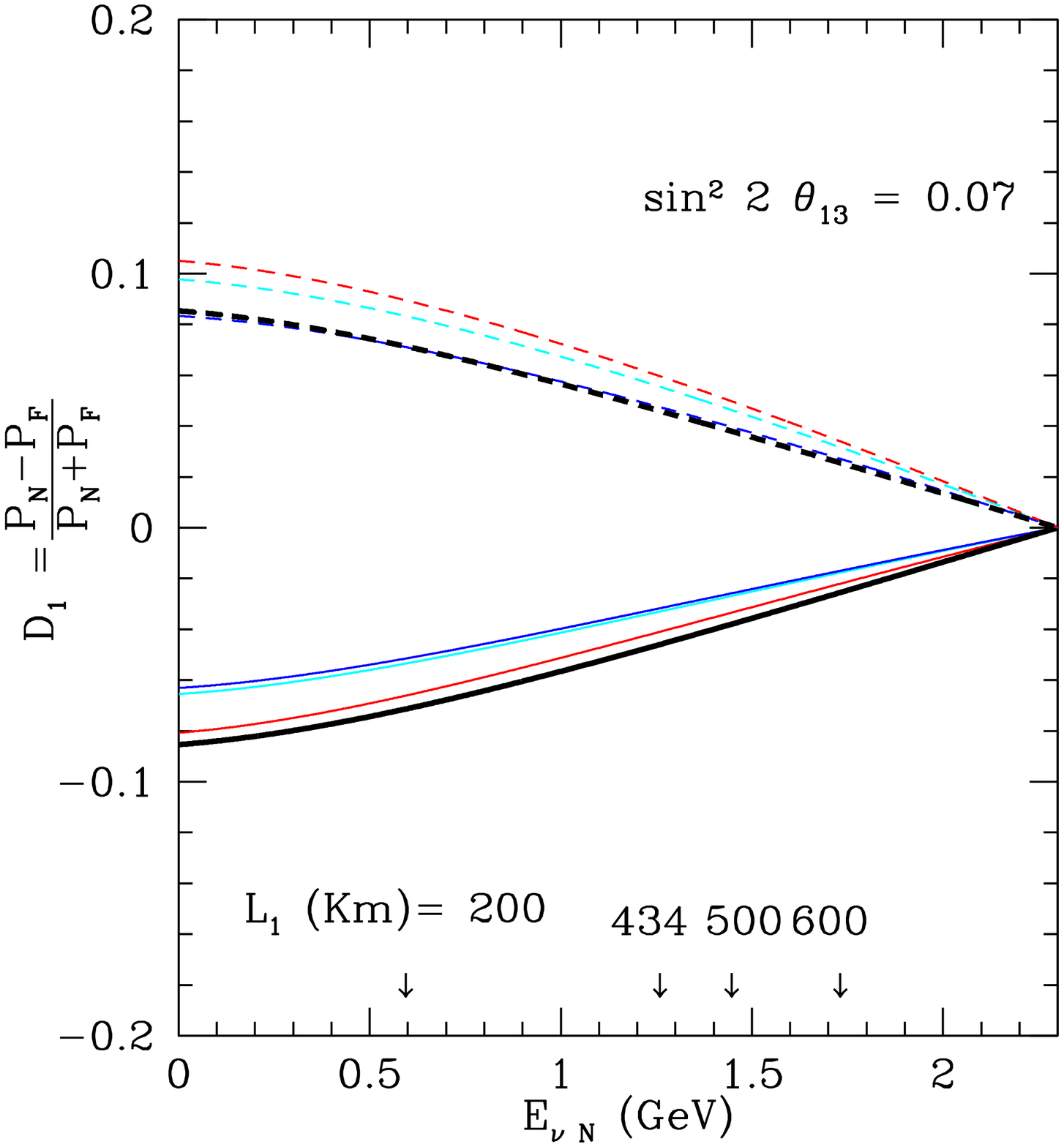, width=9.0cm} \\
\hskip 3.2truecm
{\small (a)}            &
\hskip 3.8truecm
{\small (b)}
\end{tabular}
\end{center}
\caption{\textit{(a) Approximate probability difference,
Eq.~(\ref{eq:probdiff}), as a function of the neutrino energy for
normal (black thick solid curve) and inverted hierarchies (dashed
thick black curve)  for $\sin^2 \theta_{13}= 0.02$. We also depict the
exact computation of the probability difference for the two hierarchies
and different values of $\delta$. From smaller to larger values of 
$|\probdiff|$: $\delta$ =  $\frac{\pi}{2}$ (blue), 0 (cyan) and 
$\frac{3 \pi}{2}$ (red) (color online). In the x-axis we specify the
distances that we have explored as possible locations for the near
detector, related to the neutrino energy by $L_{\rm N} = E_{\rm N}
L_{\rm F}/E_{\rm F}$. In this study we present the results for $L_{\rm
  N} = 200, 434 $ km. (b)~Same as (a) but for $\sin^2 2 \theta_{13}=
0.07$.}}
\label{fig:asymfirst}
\end{figure}

\section{Matter effects and the type of hierarchy}
\label{matter}

The first task of a long-baseline neutrino experiment should be to
measure the small $\theta_{13}$ angle of the neutrino mixing
matrix. Adding one detector to the experiment would increase the
statistics and then the sensitivity to this mixing angle. Here, we will
assume that the value of $\theta_{13}$ is within the 
sensitivity of next-generation long-baseline and reactor neutrino
experiments. We will focus on the possibility to discriminate the type
of neutrino mass ordering and will show that, by considering the
experimental setup described above, the improvement with
respect to the current \nova proposal is remarkable. The study of the
enhanced sensitivity to the value of $\theta_{13}$ and to the
CP--violating phase $\delta$ for the experimental configuration
presented here will be performed elsewhere~\cite{MPPprep}. 

\begin{figure}[t]
\begin{center}
\begin{tabular}{ll}
\epsfig{file=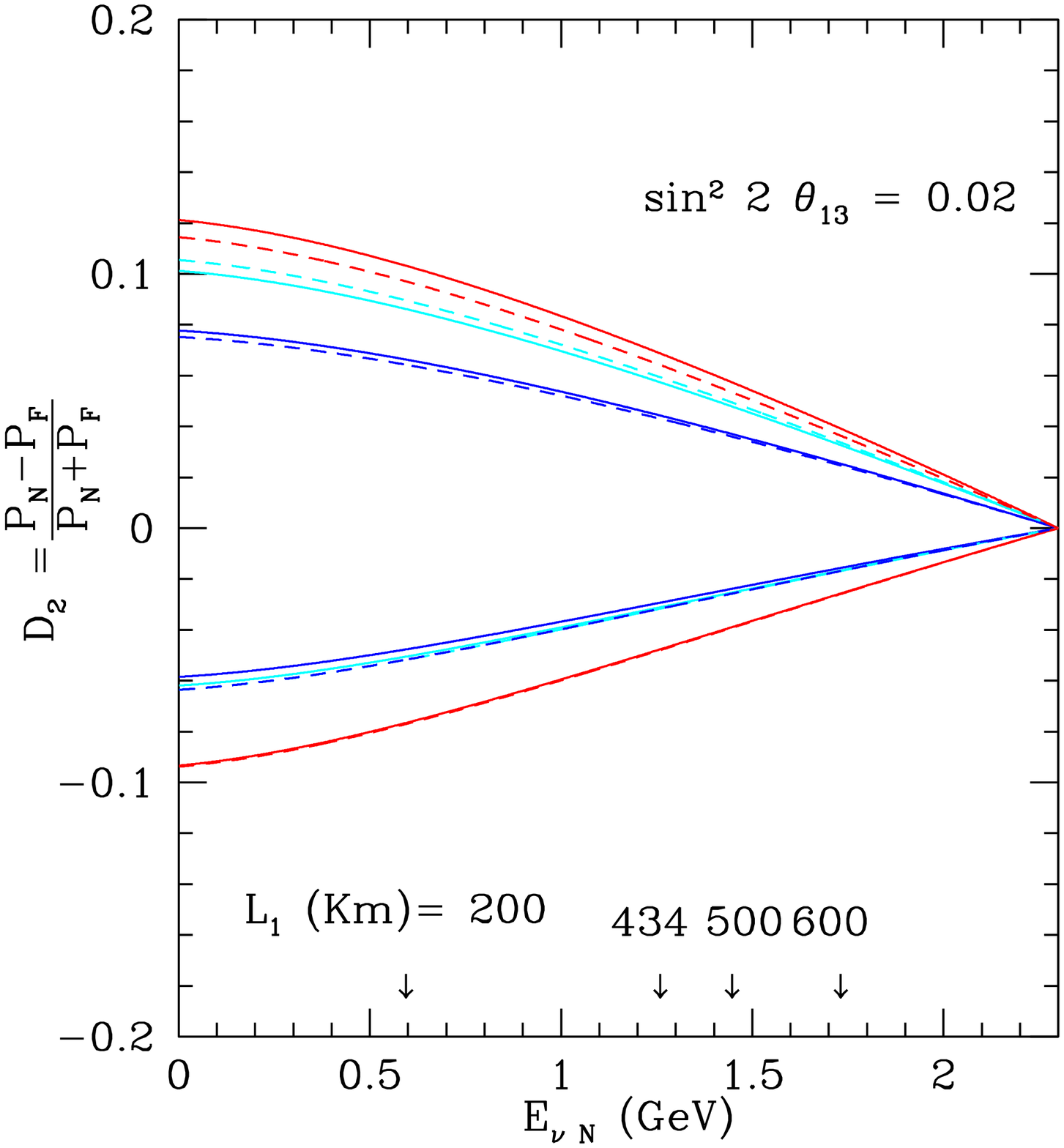, width=9.0cm} &
\hskip 0.cm
\epsfig{file=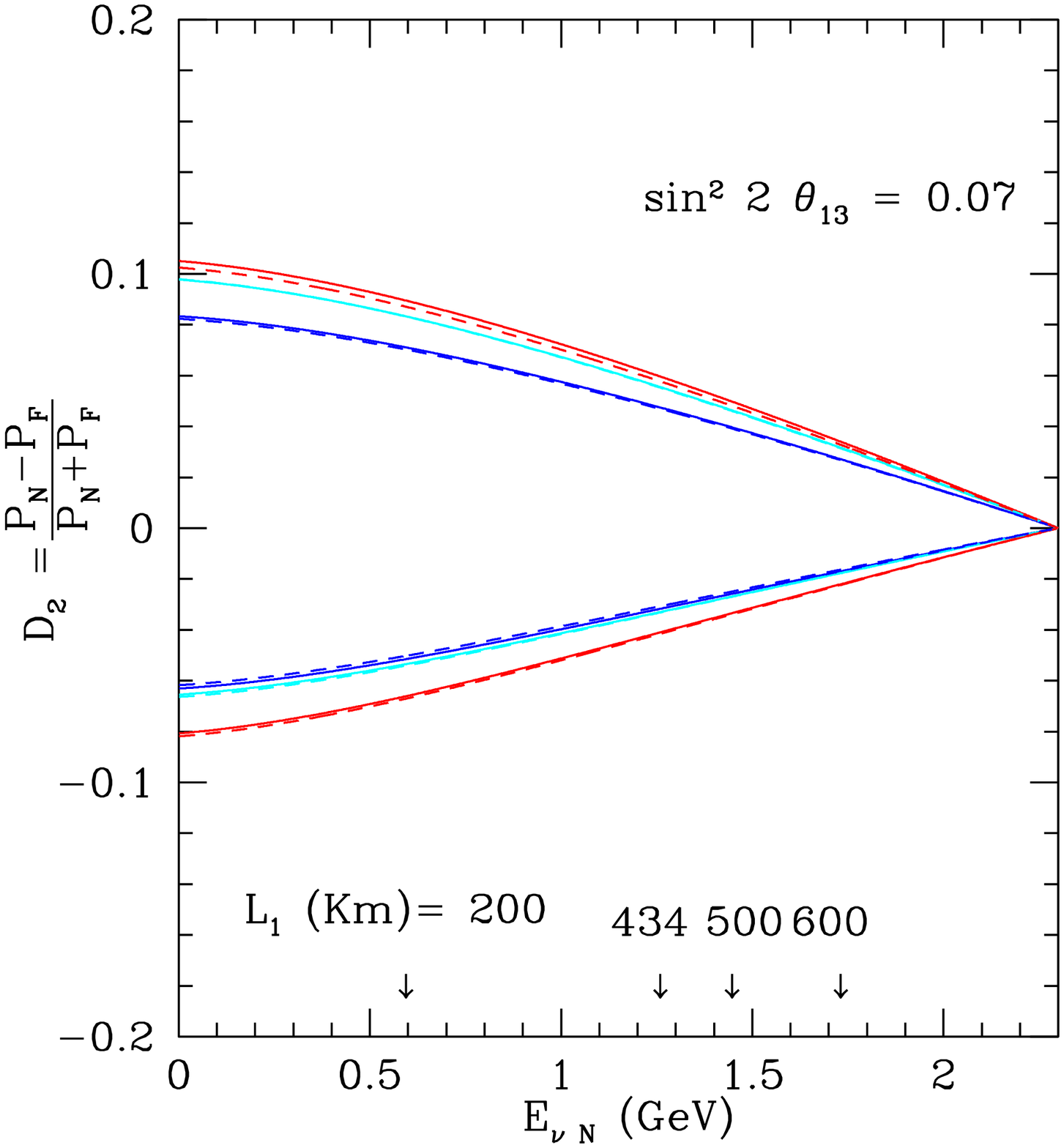, width=9.0cm} \\
\hskip 3.2truecm
{\small (a)}            &
\hskip 3.8truecm
{\small (b)}
\end{tabular}
\end{center}
\caption{\textit{(a) The dashed curves depict the probability
difference up to second order in Eq.~(\ref{eq:probdifffull}), as a
function of the neutrino energy for normal (lower curves) and inverted
hierarchies (upper curves) for $\sin^2 2 \theta_{13}= 0.02$ together
with the exact computation (solid curves). From smaller to larger
values of $|\probdiff|$, we plot three different values of $\delta$ =
$\frac{\pi}{2}$ (blue), 0 (cyan) and $\frac{3 \pi}{2}$ (red) (color
online). In the x-axis we specify the distances that we have explored
as possible locations for the near detector, related to the neutrino
energy by $L_{\rm N} = E_{\rm N} L_{\rm F}/E_{\rm F}$. In this study
we present the results for $L_{\rm N} = 200$ and $434 $ km. (b) Same
as (a) but for $\sin^2 2 \theta_{13}= 0.07$.}}    
\label{fig:asymsecond}
\end{figure}

In order to study matter effects, we will consider the probability of
$\nu_\mu$--$\nu_e$ oscillations in matter at two different lengths of
the baselines, $L_{\rm N}$ (near detector) and $L_{\rm F}$ (far
detector). And, as mentioned above, since the sensitivity to the mass
hierarchy is optimized for $L/E$ constant, we will consider both
baselines so that we keep the same ratio $L/E$ at both detectors. We
compute the quantity  
\beq
\probdiff \equiv \frac{P(L_{\rm N})-P(L_{\rm F})}{P(L_{\rm
    N})+P(L_{\rm F})} \, ,  
\label{eq:d}
\eeq
i.e. the normalized difference of the oscillation probabilities
computed at the near and far locations. By using the approximate
formula Eq.~(\ref{eq:probappr}) and expanding up to first order in
matter effects, $A L \ll 1$ and $A \ll \Delta_{13}$, neglecting the
solar neutrino mixing parameters and keeping terms of $\mathcal{O}(A \,
\theta_{13})$, $\probdiff$ reads
\begin{eqnarray}
\probdiff_1 \simeq \frac{A_{\rm N} L_{\rm N} - A_{\rm F} L_{\rm F}}{2}
  \ \left( \frac{1}{(\Delta_{13} L / 2)} - \frac{1}{\tan(\Delta_{13} L
  / 2)} \right)  \, . 
\label{eq:probdiff}
\end{eqnarray}
Let us note that the leading term in Eq.~(\ref{eq:probappr}) is
proportional to $\sin^2 (\Delta_{13} L/2)$ and cancels out in
$\probdiff_1$ because $L/E$ is the same for the near and far detector
sites. It is clear from Eq.~(\ref{eq:probdiff}) that the probability
difference $\probdiff_1$ changes sign if the neutrino mass spectrum is
normal or inverted, since it depends on the sign of
$\deltaunotre$.

The effects due to the CP--phase $\delta$, as well as those due to the
solar neutrino mixing parameters, are subleading for large enough
values of $\sin^2 \theta_{13}\geq 0.01$, region which is within the
range expected to be explored by the \nova experiment. In
Fig.~\ref{fig:asymfirst} we depict the neutrino energy dependence of
the quantity $\probdiff_1$, Eq.~(\ref{eq:probdiff}), for the two
possible hierarchies and we compare the results with the ones obtained
using the exact oscillation probabilities for three different values
of the CP--phase $\delta= 0, \frac{\pi}{2}, \frac{3 \pi}{2}$. We show
our results for two different values of $\sin^2 2 \theta_{13} = 0.02$,
close to the sensitivity limit for the \nova experiment and $\sin^2
2\theta_{13} = 0.07$, close to the present upper bound,
Eq.~(\ref{eq:chooz}).

For large enough values of $\sin^2 2 \theta_{13}$, the contribution of
CP effects to~$\probdiff_1$ is never larger than 20--30\% (see
Fig.~\ref{fig:asymfirst}). It can be shown that, at leading order, the
second term of the r.h.s. of Eq.~(\ref{eq:probappr}), which controls 
the CP--violating effects, depends only on $L/E$. Therefore, in the
normalized difference $\probdiff$, the leading CP--violating
contribution cancels out and CP--violating effects can be treated as a
perturbation to the dominant contribution from matter effects. 

The expression for $\probdiff$ with terms up to $\mathcal{O}(A \,
\Delta_{12}/\Delta_{13})$, $\mathcal{O}(A \, \theta_{13}^2)$ and
$\mathcal{O}(A^2)$ reads\footnote{In order to obtain the term
$\mathcal{O}(A \, \theta_{13}^2)$ one has to use the expression for
the probability up to $\mathcal{O}(\theta_{13}^3)$ (see
Ref.~\cite{CDGGCHMR00}), and not just Eq.~(\ref{eq:probappr}). For
large values of $\theta_{13}$ this term is of the order of the
$\mathcal{O}(A^2)$ contribution.}:   
\begin{eqnarray}
\nonumber
\probdiff_2 & \simeq  & \probdiff_1
 \left( 1- \frac{\Delta_{12}}{\Delta_{13}} \frac{\cos \theta_{13}}
   {\tan \theta_{23}} \ \frac{\sin {2 \theta_{12}}}{\sin 2
   \theta_{13}} \ \frac{\Delta_{13} L / 2}{ \sin (\Delta_{13} L / 2)}
   \cos \left( \delta + \Delta_{13} L / 2 \right) 
 - \frac{1}{2} \, \sin^2 2 \theta_{13} \nonumber \right) \\[2ex] 
 & & +
  \frac{1}{2} \frac{A_{\rm N}^2 L_{\rm N}^2 - A_{\rm F}^2 L_{\rm
      F}^2}{4} \left( 
  \frac{1}{(\Delta_{13} L / 2)^2} - \frac{1}{\sin^2 \left( \Delta_{13}
   L / 2 \right)} \right)~, 
\label{eq:probdifffull}
\end{eqnarray}
where the correction due to the CP--phase also changes sign  
with the hierarchy type. For instance, for normal hierarchy, the
correction to $\probdiff_1$ due to the first term in the r.h.s. of
Eq.~(\ref{eq:probdifffull}) will increase $|\probdiff_1|$ if $\pi/2 <
\delta + |\Delta_{13}| L / 2 < 3 \pi/2$ whereas for inverted hierarchy
it will decrease $|\probdiff_1|$ if $ \pi/2 < \delta - |\Delta_{13}| L
/ 2 < 3 \pi/2$. However, the second-order correction due to matter
effects does not depend on $sign(\deltaunotre)$ and it is always
negative.

\begin{figure}[t]
\begin{center}
\begin{tabular}{ll}
\epsfig{file=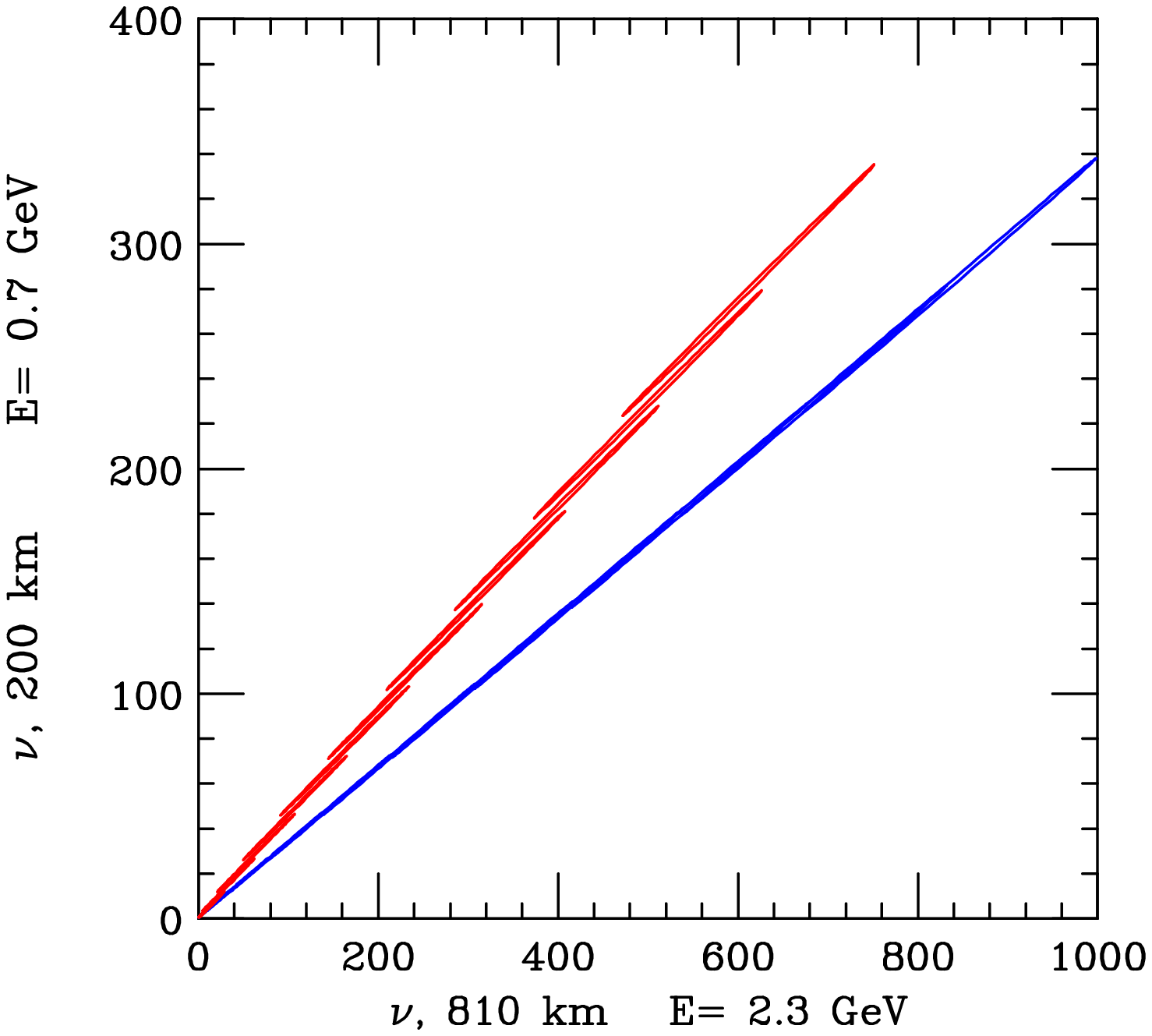, width=9.0cm} &
\hskip 0.cm
\epsfig{file=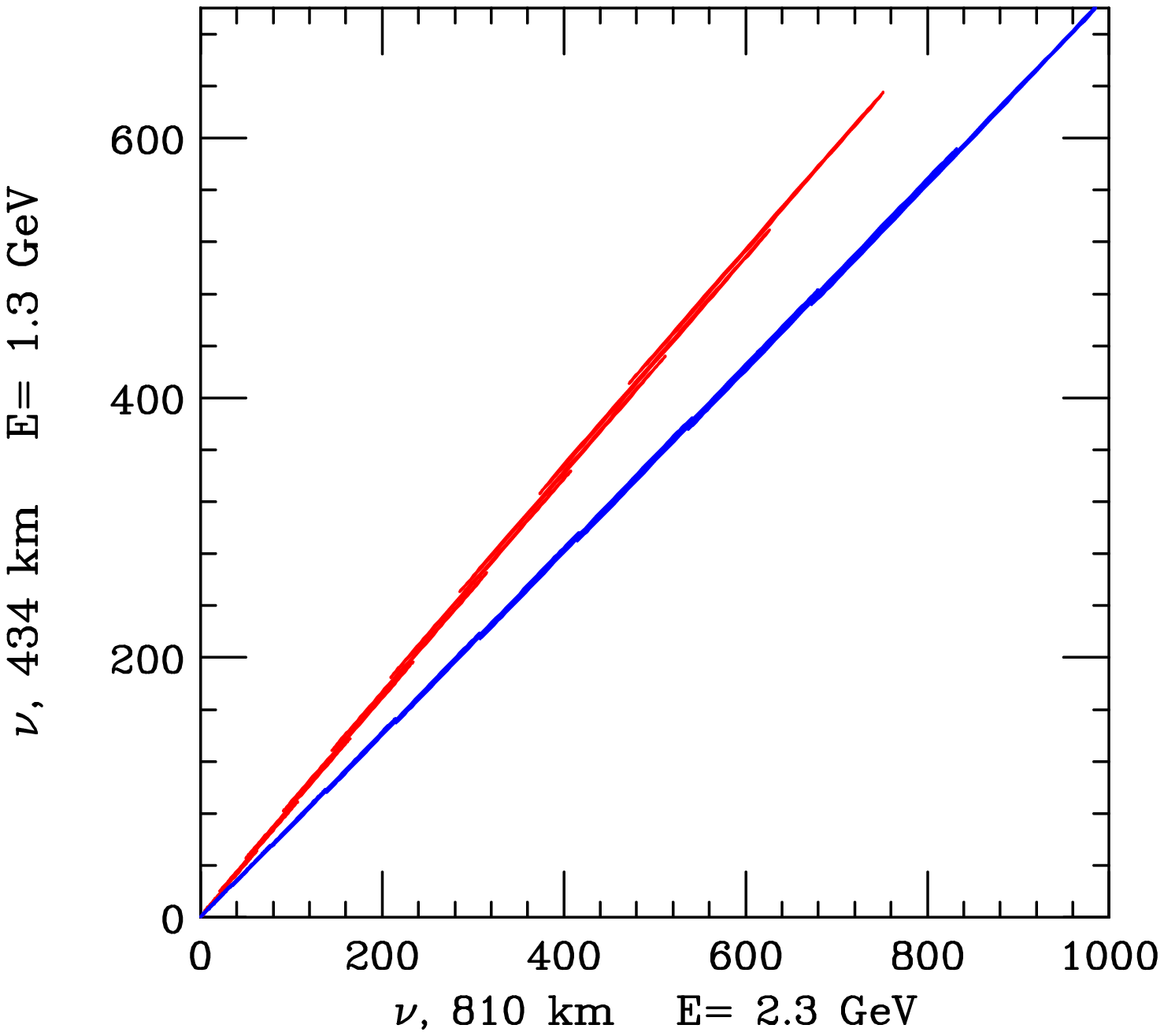, width=9.0cm} \\
\hskip 3.2truecm
{\small (a)}            &
\hskip 3.8truecm
{\small (b)}
\end{tabular}
\end{center}
\caption{\textit{(a) Bi--neutrino event ellipses at the short (200 km)
and at the far distances (810 km) for normal (lower blue) and inverted
(upper red) hierarchies. The experimental setup considered here is
$3.7 \times 10^{20}$ protons on target per year, a 50 kton detector
with perfect efficiencies at each location and five years of data
taking. From bottom up, the ellipses correspond to $\sin^2 2
\theta_{13}= 0.0003, 0.0004, 0.006, 0.008, 0.001, 0.005, 0.01, 0.02,
0.03, 0.04, 0.06, 0.07, 0.095$ and $0.11$. (b) Same as (a) but with
the near detector located at 434 km.}}  
\label{fig:binu}
\end{figure}

The probability difference up to second order,
Eq.~(\ref{eq:probdifffull}), is depicted in Fig.~\ref{fig:asymsecond}  
for the three different values of the CP--phase $\delta=0,
\frac{\pi}{2}, \frac{3 \pi}{2}$ and for the two possible
hierarchies. We repeat the same exercise as in
Fig.~\ref{fig:asymfirst} and show our results for two different values
of $\theta_{13}$: $\sin^2 2\theta_{13}  = 0.02$ and $\sin^2 2
\theta_{13} = 0.07$.  

We have thus shown that, under this special experimental
configuration, determining the type of hierarchy requires only
establishing whether $\probdiff$ is positive or negative, being this
measurement free of other degeneracies; the corrections due to the
rest of the neutrino mixing parameters cannot flip this sign. This
implies that the determination of the type of hierarchy by using only
the neutrino channel with two detectors at different baselines suffers
no degeneracies with other parameters. In addition, the requirement of
just only the neutrino channel (with two detectors, though) will allow
the number of years of data taking (and systematic uncertainties) to
be reduced.

In order to illustrate the potential of the combination of two
long-baseline detectors operating at the same $L/E$ in extracting
$sign(\deltaunotre)$, we show here the neutrino bi--event plots at the 
far and at the short distance detectors and we compare these results
with the neutrino--antineutrino bi--event plots at just one fixed
distance. In Fig.~\ref{fig:binu} we show the neutrino bi--event curves
for the two hierarchies at the short and at the far distance,
considering two short baselines (200 km and 434 km) and a fixed
long-baseline  (810 km) for different values of $\sin^2 2
\theta_{13}$. As it is clearly seen from these plots, for the case of
constant $L/E$ at both detectors, the ellipses collapse to a line and
those obtained if the solution is that of $sign(\deltaunotre) > 0$ no
longer overlap with those for $sign(\deltaunotre) < 0$. Notice also
that the slope for the normal hierarchy ``pencil'' is smaller than
that for the inverted hierarchy, because of the larger matter effect
for larger baselines (for neutrinos).

\begin{figure}[t]
\begin{center}
\epsfig{file=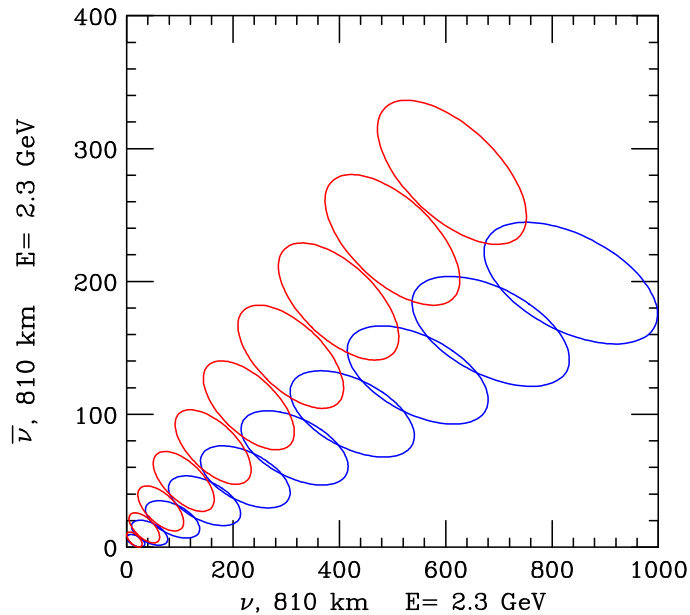,width=9.0cm} 
\caption{\textit{Bi--event neutrino--antineutrino ellipses at the far
distances (810 km) for normal (lower blue) and inverted (upper red)
hierarchies. The experimental setup is the one described in
Fig.~\ref{fig:binu}, but considering 5 years of neutrino running and
5 years of antineutrino running. From bottom up, the ellipses
correspond to $\sin^2 2 \theta_{13}=0.001, 0.005, 0.01, 0.02, 0.03,
0.04, 0.06, 0.07, 0.095$ and  $0.11$.}} 
\label{fig:bianu}
\end{center}
\end{figure}

If instead we consider the commonly assumed configuration of using a
single detector, and running first in the neutrino mode and then in the
antineutrino mode, the ellipses overlap for a large fraction of values 
of the CP--phase $\delta$ for every allowed value of $\sin^2 2
\theta_{13}$. This makes the determination of $sign(\deltaunotre)$ 
extremely difficult, i.~e., the $sign(\deltaunotre)$-extraction is not
free of degeneracies. The former case is depicted in
Fig.~\ref{fig:bianu}, where we have considered 5 years of neutrino and 
antineutrino running. Therefore, even with half of the time of data
taking, placing two detectors could resolve the type of neutrino mass
hierarchy much more easily than the standard approach. 

In what follows, we will provide a detailed study of the sensitivity
to the $sign(\deltaunotre)$ in the $\sin^2 2 \theta_{13}$--$\delta$
plane.

In order to compute the sensitivity to the mass hierarchy, as a first
approach, we have constructed a measurable integrated asymmetry:
\beq
A_{+} = \frac{ \{ N / N_o \}_{\rm N} 
                            - \{ N / N_o  \}_{\rm F}}{
                               \{ N / N_o \}_{\rm N} 
                             + \{ N / N_o \}_{\rm F}} \, , 
\label{intasy}
\eeq
where $N$ is the number of $\nu_e$ induced events in the presence of
oscillations in the normal hierarchy scheme and $N_o$ is the expected
number of $\nu_\mu$ charged-current interactions in the absence of
oscillations at the near (N) and far (F) detectors. One can compute
the equivalent integrated asymmetry assuming an inverted hierarchical
scenario, $A_{-}$. If Nature has chosen, for instance, a positive
value for the atmospheric splitting but the data analysis is performed
assuming the opposite sign, the sensitivity to the sign resolution reads
\beq
\frac{ |A_{+} -A_{-}|}{ \delta {A}_{+} } \, ,
\label{eq:stat}
\eeq
i.~e., the difference between the two asymmetries divided by its error. 
We have studied the case of statistical errors (adding backgrounds) as
well the impact of the uncertainties on the remaining oscillation
parameters by computing this \emph{systematic error} on the asymmetry
using the standard error propagation method. The errors on the solar
parameters can be safely neglected. The errors considered here for the
atmospheric mixing parameters $\deltaunotre$ and $\sin^2 2
\theta_{23}$ are at the level of $5\%$ and $2\%$
respectively~\cite{NOvA,messierp04}. For the matter parameter we take 
the conservative assumption $\Delta A/A=5\%$~\cite{refmatter}. 

\begin{figure}[t]
\begin{center}
\begin{tabular}{ll}
\epsfig{file=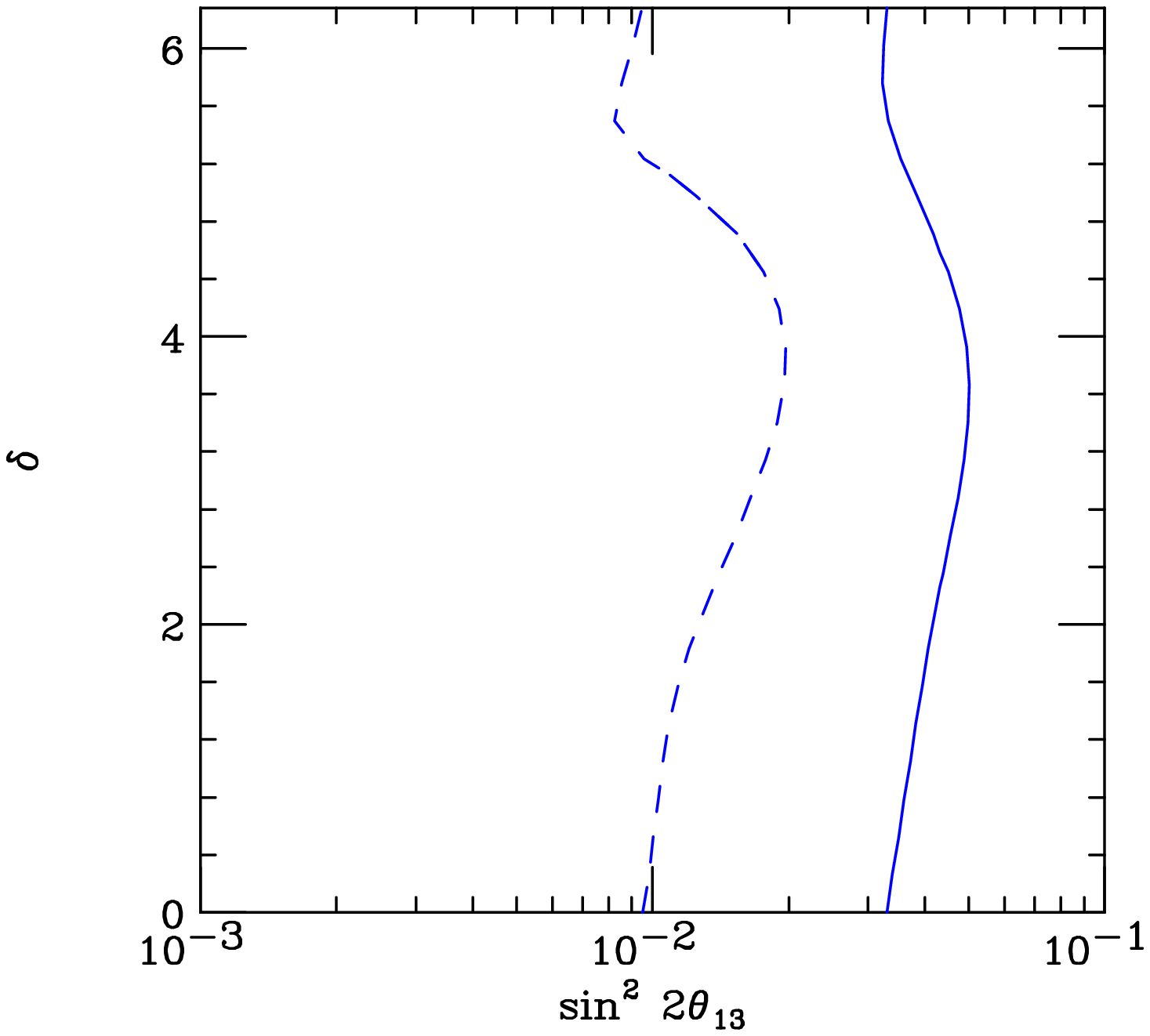, width=9.0cm} &
\hskip 0.cm
\epsfig{file=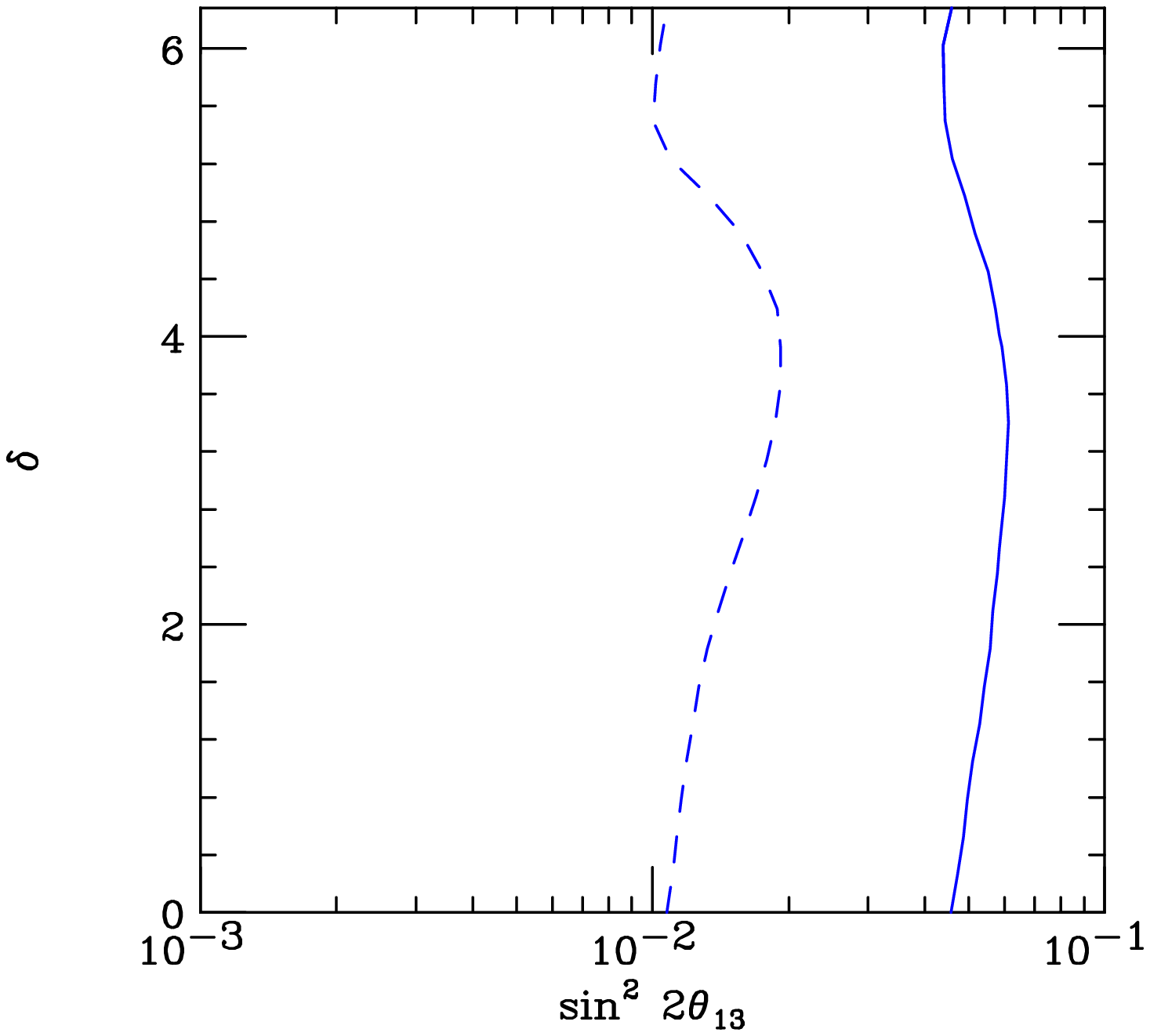, width=9.0cm} \\
\hskip 3.2truecm
{\small (a)}            &
\hskip 3.8truecm
{\small (b)}
\end{tabular}
\end{center}
\caption{\textit{(a) Sensitivity to the sign of the atmospheric mass
square difference (for $|\deltaunotre| = 2.4 \times 10^{-3} \
\rm{eV}^2$) as defined in Eq.~(\ref{eq:stat}), including the
systematic errors induced by the uncertainties on the atmospheric
mixing parameters and on the matter parameter $A$, exploiting the data
from a short-baseline off-axis detector located at 200 km and from the 
long-baseline off-axis detector at 810 km. The experimental setup
assumed to obtain the solid curves is $3.7 \times 10^{20}$ protons on
target per year, a 50 kton detector with perfect efficiencies at each
location and five years of data taking, whereas to get the dashed
curves we have added to the former statistics a factor of 5 (proton
driver case). (b) Same as (a) but combining the data from the
long-baseline with a short-baseline off-axis experiment located at 434
km.}}   
\label{fig:sens}
\end{figure}

We present the $95\%$ C.L. sensitivities to the neutrino mass
hierarchy resolution in Fig.~\ref{fig:sens}, where Nature's solution  
for the neutrino mass spectrum has been chosen to be the normal
hierarchy. We have studied both possibilities, normal and inverted
hierarchy, as Nature's or ``true'' solution. We find that the
conclusions do not change in a significant way when considering the 
inverted neutrino mass spectrum. 

Two possible combinations of the data at the far, ``fixed'' $\nova$
off-axis experiment have been explored: with the data from a near
detector located at 200 km and with those from a near detector
located at 434 km. All these experiments would exploit the NuMI
neutrino beam in an off-axis mode. From the results depicted in
Fig.~\ref{fig:sens}, one can observe that the best option for the
location of the second, near detector, would be 200 km: the hierarchy
could be determined regardless of the value of the CP--phase $\delta$
down to values of $\sin^2 2 \theta_{13}=0.05$ or $\sin^2 2
\theta_{13}=0.02$ with the Proton Driver option. For a longer
baseline, 434 km, the type of hierarchy can be uniquely determined
independently of $\delta$ for $\sin^2 2 \theta_{13}=0.06$ and $\sin^2
2 \theta_{13}=0.02$, without and with the Proton Driver, respectively. 

For comparison, we consider the study performed in the revised $\nova$
proposal~\cite{newNOvA} on the sensitivity to the mass hierarchy. The
analysis considers 3 years of running in the neutrino and antineutrino
modes, with a 30~kton detector located at a baseline of 820 km and 12
km off-axis. A value of $\deltaunotre = 2.5 \times 10^{-3}\ {\rm eV}^2$
was chosen. $\nova$ alone can resolve the sign of $\deltaunotre$ at the
$95\%$ C.L. for only $10\%$ of the values within the range of the
CP--phase $\delta$ if $\sin^2 2 \theta_{13}=0.04$. Even if $\sin^2 2
\theta_{13}=0.1$, a value that is very close to the present upper
bound, \nova could uniquely determine the type of hierarchy for only
$40\%$ of the values of the CP--phase $\delta$~\cite{newNOvA}. Even
with the Proton Driver, $\nova$ cannot resolve the sign of the
atmospheric mass difference for 80$\%$ (40$\%$) of the values of
$\delta$ if $\sin^2 2 \theta_{13} \le 0.02 (0.1)$~\cite{newNOvA}. 
In addition, the possibility of adding a second 50~kton detector at
$\sim 700$~km and 30 km off-axis, is discussed~\cite{cooper,newNOvA}.
In this case, the resolution of the mass hierarchy down to $\sin^2 2
\theta_{13} = 0.02$ can be accomplished after 12 years of \nova data
plus 6 years with the second detector equally split between neutrinos
and antineutrinos and with Proton Driver. This is comparable to what
the Super-NO$\nu$A setup proposed here can achieve after 5 years of
running with only neutrinos and with Proton Driver (see
Fig.~\ref{fig:sens}). However, a direct comparison needs to take into
account the different choices of detectors, the numbers of protons on
target, the value of $\deltaunotre$ used in the analysis, and the fact
that an optimization of the experiment has not been  performed in our
case. 

\begin{figure}[t]
\begin{center}
\begin{tabular}{ll}
\hskip -0.5cm
\epsfig{file=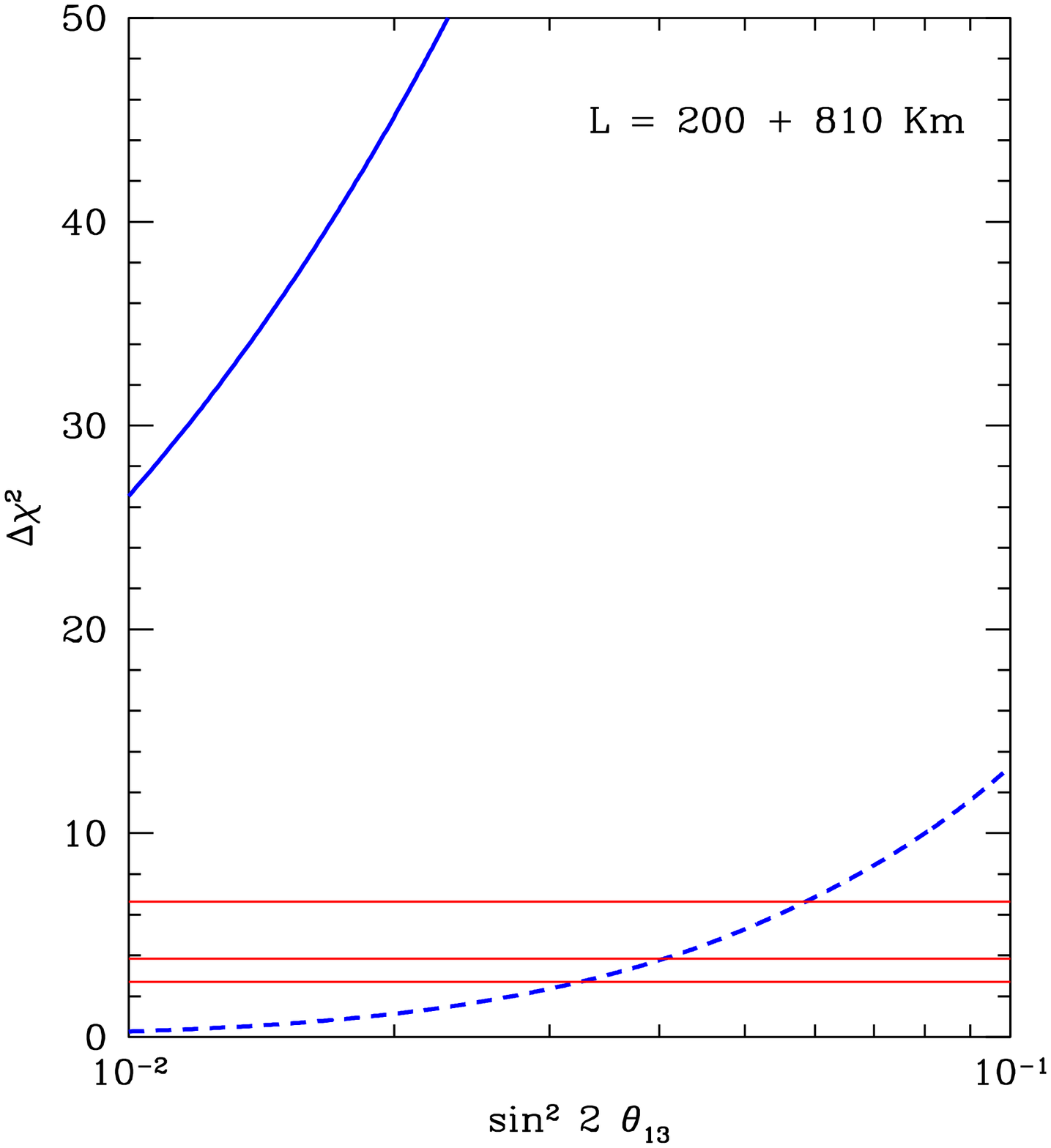, width=8.1cm} &
\hskip 0.cm
\epsfig{file=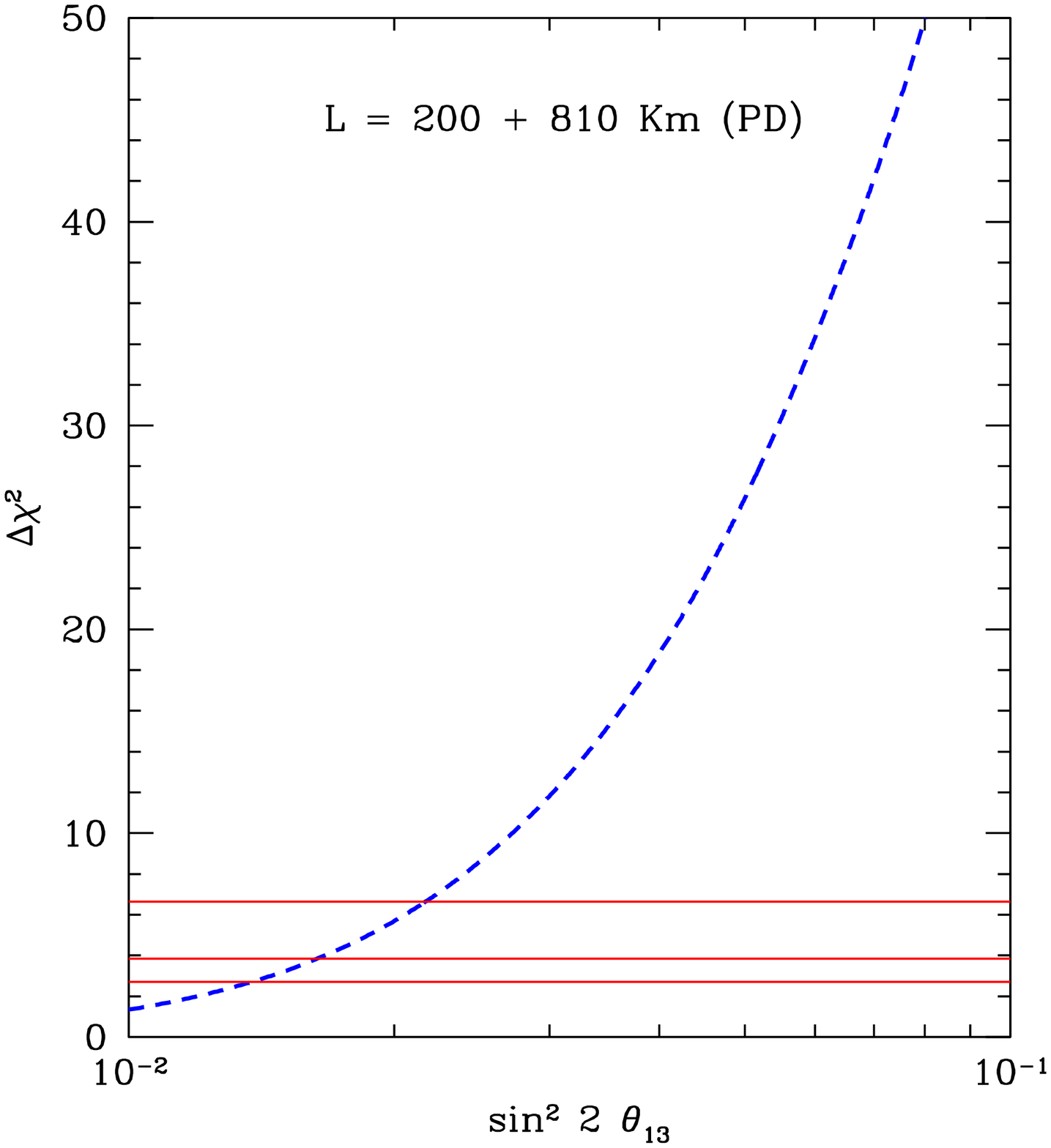, width=8.1cm} \\
\hskip 3.2truecm
{\small (a)}            &
\hskip 3.8truecm
{\small (b)}
\end{tabular}
\end{center}
\caption{\textit{(a) Results of the $\chi^2$ analysis to the sign of
the atmospheric mass difference extraction versus $\sin^2 2
\theta_{13}$, by exploiting the data from a far long-baseline
experiment at 810 km and from a short-baseline experiment at 200 km,
for $|\deltaunotre| = 2.4 \times 10^{-3} \ \rm{eV}^2$. The
corresponding $90\%$, $95\%$ and $99\%$ C.L.s are shown. As a function
of $\sin^2 2 \theta_{13}$, we depict the maximum (solid line) and
minimum (dashed line) of $\Delta \chi^{2}$, which are obtained for
different values of $\delta$ depending on $\sin^2 2 \theta_{13}$. (b)
Same as (a) but with a Proton Driver.}}
\label{fig:chi2} 
\end{figure}

We have performed an independent $\chi^{2}$ analysis of the data on the
$\sin^{2} 2 \theta_{13}$--$\delta$ plane. At a fixed value for the 
former two parameters, the $\chi^2$ in the combination of two
baselines (near and far sites) reads
\beq
\chi_{\ell \ell'}^2 = \sum_{\ell \ell'} \; (\mathcal{N}_{\ell,\pm} -
N_{\ell,\pm}) C_{\ell:\ell'}^{-1} (\mathcal{N}_{\ell',\pm} -
N_{\ell',\pm})\,, 
\label{chi2c}
\eeq
where the + ($-$) sign  refers to normal (inverted) hierarchy and $C$
is the covariance matrix, which for the particular analysis considered
in the present study contains only statistical errors. The
experimental ``data'', ${\mathcal N_{\ell,\pm}}$, are given by   
\beq
\mathcal N_{\ell,\pm} = \langle N_{\ell,\pm} + N_{b\ell }\rangle -
N_{b\ell, \pm}\, , 
\eeq
where we have considered that the efficiencies are flat in the visible
energy window, $N_{b\ell }$ are the background events and $\langle
\rangle$ means a Gaussian/Poisson smearing (according to the
statistics). We have assumed that nature has chosen a given sign for
$\deltaunotre$, but the data analysis is performed with the opposite
sign. We show the results of our $\chi^2$ analysis in
Fig.~\ref{fig:chi2}, where we plot the sensitivity to
$sign(\deltaunotre)$ after the combination of the data from the
long-baseline (810~km) with the data from a short baseline at 200
km. For every value of $\sin^2 2 \theta_{13}$ we find the two
values\footnote{The values of $\delta$ at the maximum and minimum 
of $\Delta \chi^2$ are in general different for different values of
$\sin^2 2 \theta_{13}$.} of $\delta$ that maximize and minimize
$\Delta \chi^2$. We depict in Fig.~\ref{fig:chi2} the value of the
minimum and maximum $\Delta \chi^2$ versus $\sin^2 2 \theta_{13}$. If
$\sin^2 2 \theta_{13}\ge 0.04$ a misidentification in the sign of the
atmospheric mass difference can be excluded at $95\%$ C.L. in the most
pessimistic situation. There exists, however, a large number of values
of $\delta$ at which this misidentification can be excluded at a
confidence level larger that $99\%$. With a proton driver, if $\sin^2
2 \theta_{13} \ge 0.02$, our analysis shows that it is possible to
determine $sign(\deltaunotre)$ at the level of the $99\%$ C.L. for the
full range of $\delta$. The results from the $\chi^2$ analysis
presented here agree with the previous study of the asymmetries.

\subsection{Dependence on  $|\deltaunotre|$}

\begin{figure}[t]
\begin{center}
\epsfig{file=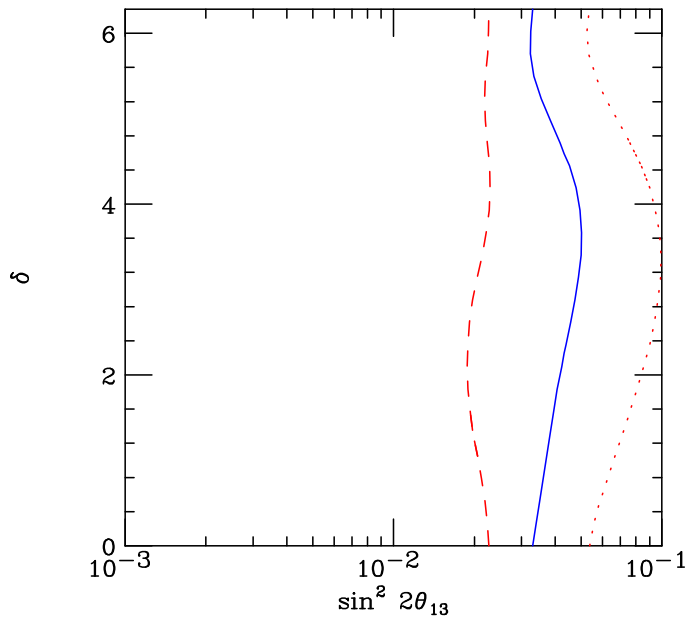,width=9.0cm} 
\caption{\textit{Sensitivity to the sign of the atmospheric mass
square difference as defined in Eq.~(\ref{eq:stat}), for different
values of $|\deltaunotre|$ = $2.0 \times 10^{-3} \ \rm{eV}^2$ (dotted
red line), $2.4 \times 10^{-3} \ \rm{eV}^2$ (solid blue line), and $3.0
\times 10^{-3} \ \rm{eV}^2$ (dashed red line). We have included the
systematic errors induced by the uncertainties on the atmospheric
mixing parameters and on the matter parameter A, and  exploited the
data from a short-baseline off-axis detector located at 200 km and
from the long-baseline off-axis detector at 810~km. The experimental
setup assumed to perform the solid curves is $3.7 \times 10^{20}$
protons on target per year, a 50 kton detector with perfect
efficiencies at each location, and five years of data taking.}} 
\label{fig:difdm}
\end{center}
\end{figure}

In the previous section, we have assumed the knowledge of
$|\deltaunotre|$ with a 5\% uncertainty. In particular, we have taken
$|\deltaunotre| = (2.40 \pm 0.12 ) \times 10^{-3} \
\rm{eV}^2$. Although the former level of precision is expected to be
achieved by the time this experiment could turn
on~\cite{NOvA,messierp04}, currently the value of the atmospheric mass
difference is not known with that level of accuracy. Thus, it is
important to investigate how the results presented above change if
$|\deltaunotre|$ happens to be different from the previously assumed
value.

In Fig.~\ref{fig:difdm} we have depicted the sensitivity to the sign
of the atmospheric mass square difference determination as defined in
Eq.~(\ref{eq:stat}), for three different possibilities for its
absolute value, $|\deltaunotre|$ = $2.0
\times 10^{-3} \rm{eV}^2$ (dotted red line); $2.4 \times 10^{-3}
\rm{eV}^2$ (solid blue line) and $3.0 \times 10^{-3} \rm{eV}^2$
(dashed red line). We have assumed a near off-axis detector located at
200 km. As can be seen from the figure, the larger the value of
$|\deltaunotre|$, the better the sensitivity to the type of neutrino
mass hierarchy. The reason can be easily understood from
Eq.~(\ref{eq:probdiff}). The asymmetry depends on the factor $1/x -
1/\tan x$, where $x \propto \deltaunotre$ (recall that $L/E$ is the
same for both detectors). This function increases monotonically as $x$
increases, and therefore the asymmetry is larger for larger
$|\deltaunotre|$, which correspondingly means better sensitivity. 

On the other hand, it must be remarked that since the
CHOOZ~\cite{CHOOZ} bound is weaker for small values of 
$|\deltaunotre|$, the loss in range for $\theta_{13}$ is not as large
as one would na\"{\i}vely think from Fig.~\ref{fig:difdm}.

In case the actual value of $|\deltaunotre|$ happens to be in the low
side of the currently allowed range~\cite{SKatm}, a possible solution
could be to adopt a larger $L/E$, which can be accomplished either by
considering longer baselines or larger off-axis angles, i.e., smaller
energies. However, both possibilities imply the reduction of the
flux at the detectors, so a compromise must be achieved. Nevertheless,
if $\theta_{13}$ is very small and $|\deltaunotre|$ is also small,
then a longer run in the neutrino mode would unavoidably be needed in
order to increase the statistics. All in all, a detailed analysis
would be required to find the best possible configuration as a
function of $|\deltaunotre|$~\cite{MPPprep}.

\section{Conclusions}
\label{conclusions}
Establishing the type of neutrino mass hierarchy --- be it
normal or inverted --- plays a crucial role in our understanding of
neutrino physics. Future long-baseline experiments will address this
fundamental question. Typically the determination of the hierarchy in
the proposed experiments suffers from degeneracies with other
CP--conserving and CP--violating parameters, namely $\theta_{13}$,
$\delta$ and $\theta_{23}$. Resolving such degeneracies in one
experiment is very challenging, if not impossible. Different strategies
have been studied, e.~g., by combining more than one
experiment~\cite{BMW02,HLW02,MNP03,otherexp,mp2,unveiling}, using more
than one detector~\cite{BNL,MN97,BCGGCHM01,silver,BMW02off}, or using
additional information from atmospheric neutrino data~\cite{HMS05}. 

In the present article, we have presented a method for establishing
$sign(\deltaunotre)$, free of degeneracies, by using only one
experiment running in the neutrino mode alone. We have considered an
experimental setup with two neutrino detectors placed in a special
off-axis configuration. It is known that off-axis spectra are well
peaked at a certain neutrino energy, which depends on the angle from
the central axis of the beam. This allows  both detectors to be
located in such a way that they have the same $L/E$. We have shown
that very interesting synergy effects show up with this special
configuration for which vacuum oscillation phases are the same at both
sites, stressing the different matter effects. These are manifest when
comparing the bi--event neutrino--neutrino (Fig.~\ref{fig:binu}) and
bi--event neutrino--antineutrino (Fig.~\ref{fig:bianu}) plots above,
for which a clear distinction of the type of hierarchy is possible for
the former regardless of the value of $\delta$, but more challenging
for the latter. 

We have considered a normalized difference, $\probdiff$ (see
Eq.~(\ref{eq:d})), between the neutrino oscillation probabilities at
two baselines. At leading order, the sign of $\probdiff$ depends only
on matter effects, i.~e., on the type of neutrino mass hierarchy, while
other parameters are subdominant. Although CP--violating terms can have
a sizable contribution for small values of $\sin^2 2\theta_{13}$, they
cannot change the sign of $\probdiff$. This implies that the
determination of the type of hierarchy, exploiting the method
discussed here, suffers no degeneracies from other parameters. We have 
confirmed such a result by performing an analysis of the sensitivity to
the mass hierarchy in a specific experimental setup, which we have
named Super-NO$\nu$A. We propose to use the NuMI beam in the neutrino
mode and two detectors, one at the far distance proposed by the \nova
Collaboration, $L_{\rm F} \sim 800$~km, and the other with a shorter
baseline, $L_{\rm N} \sim 200$~km (434 km), with the energy tunned to
$E_{\rm N} = E_{\rm F} L_{\rm N}/L_{\rm F}$. The selection of the
short baseline must be done in such a way that it is possible to place
a detector at the precise off-axis angle, in order to get that
particular energy at the peak of the spectrum. Because of the Earth
curvature, the near detector, located on the Earth surface and on the
vertical of the on-axis beam, is off-axis by a small angle, which is
the minimum possible off-axis angle at that distance. This implies
that not all different configurations, such that $L/E$ is the same at
both sites, are possible. In particular, there are no sites between
300 and 400 km which give an $L/E$ ratio of 352 km/GeV.

By considering the integrated asymmetry, Eq.~(\ref{intasy}), we have
shown that a suitable baseline for the second detector to determine
the type of neutrino mass hierarchy corresponds to $L_{\rm N} \sim
200$~km, which enhances matter effects without the need of too low
energies to keep the same $L/E$ at both detectors. We have shown in
Figs.~\ref{fig:sens} and~\ref{fig:chi2} that this can be achieved at 
95\%~C.L., regardless of the value of $\delta$, for $\sin^2
2\theta_{13}\geq 0.05$ for a conventional beam and for $\sin^2
2\theta_{13}\geq 0.02$ with a Proton Driver. Similar results can be
obtained for a slightly longer baseline, e.~g., 434~km. We have also
performed an independent $\chi^{2}$ analysis of the simulated data on
the  $\sin^{2} 2 \theta_{13}$--$\delta$ plane, which confirms
our previous results. This is in contrast with the sensitivity of the
proposed \nova experiment. At the $95\%$ C.L., only for $10\%$ of the
values of the CP--phase $\delta$, \nova can resolve the type of
neutrino mass hierarchy if $\sin^2 2 \theta_{13}=0.04$, considering
three years of neutrino plus three years of antineutrino
running~\cite{NOvA,newNOvA}.  

In Fig.~\ref{fig:difdm} we have also shown the sensitivity to the type
of neutrino mass hierarchy for three values of $|\deltaunotre|$ for
the adopted configuration, and as can be seen from it, if
$|\deltaunotre|$ lies in the low side of the presently allowed
range~\cite{SKatm}, another configuration~\cite{MPPprep} or more
statistics might be needed.   

For our simulations we have considered two 50 kton liquid argon
detectors. In addition to the off-axis experiment detailed here (and
in general, to any long-baseline neutrino experiment), the physics
potential of liquid argon detectors is remarkable. They can also be
used as supernova neutrino detectors, for proton-decay searches, and
for studies of neutrinoless double beta decay (see Ref.~\cite{flare}
and references therein).  

It is important to notice that the use of the neutrino mode alone
would allow a reduction of the number of years of data taking if
compared with the standard approach of running in the neutrino and
then antineutrino modes. In addition, having two identical detectors
and only one beam reduces the systematic uncertainties.   

Thus, we have presented an improved off-axis experiment with respect
to the proposed \nova experiment with a high sensitivity to the type
of neutrino mass hierarchy (free of degeneracies) even with only a
neutrino run. The improved capabilities for measuring the value of
$\theta_{13}$ and the CP--violating phase, as well as other possible
configurations, will be studied elsewhere~\cite{MPPprep}.

\section{Acknowledgments}
We are indebted to Mark Messier and Adam Para for providing us with
information about the off-axis neutrino fluxes and to Andr\'e de
Gouv\^ea and Stefano Rigolin for a careful reading of the
manuscript. OM would like to thank Stephen Parke for enlightening
suggestions about this work. SP would also like to thank Maurizio Piai
for stimulating discussions at the early stages of this work. Our
calculations made extensive use of the Fermilab General-Purpose
Computing Farms~\cite{farms}. SPR is supported by NASA Grant
ATP02-0000-0151 and by the Spanish Grant FPA2002-00612 of the
MCT. Fermilab is operated by URA under DOE contract DE-AC02-76CH03000.

\appendix 

\section{}
\label{appendix}

We present in this appendix the computed charged-current neutrino
event rates for the NuMI beam and different locations of a 50 kton
liquid argon detector. As a matter of illustration, we show these
event rates in Tables~\ref{tab:nova800},~\ref{tab:nova400}
and~\ref{tab:nova200} for a given value of $\sin^2 2\theta_{13} =
0.058$, a given value of $|\deltaunotre| = 2.4 \times 10^{-3}
 \ \rm{eV}^2$,
 and for four different values of the CP--phase, $\delta = 0,
\pi/2, \pi, 3\pi/2$. In these tables we show the unoscillated
$\nu_\mu$-like events, the expected oscillated $\nu_e$-like signal and
the $\nu_e$ intrinsic background, which mainly comes from $\mu$ decays.

\begin{table}[h]
\vspace{1cm}
\centering
\begin{tabular}{|c|c|c|c|c|c|}
\hline\hline
$\sin^{2} 2 \tetaot$ & $\delta$ & $\Delta m_{31}^2$ (eV$^2$)  &
$\nu_\mu$(unoscillated) &$\nu_e$ (signal) & $\nu_e$ (intrinsic
background)\\    
\hline\hline
0.058 & 0 & 0.0024& 16322 &481 &59\\
0.058 & $\pi$/2 & 0.0024& 16322 & 526 &59\\
0.058 & $\pi$ & 0.0024& 16322 &367 &59\\
0.058 & 3$\pi$/2 & 0.0024& 16322 & 323 &59\\
\hline\hline 
0.058 & 0 & $-0.0024$& 16322 &264 &59\\
0.058 & $\pi$/2 &$ -0.0024$& 16322 & 398 &59\\
0.058 & $\pi$ & $-0.0024$& 16322 &353 &59\\
0.058 & $3\pi$/2 & $-0.0024$& 16322 & 219 &59\\
\hline\hline 
\end{tabular}
\caption{\it 
Calculated charged-current neutrino event rates (signal and
backgrounds) for NO$\nu$A (baseline of 810 km, 10 km off-axis), for 
$3.7 \times 10^{20}$ pot/yr, 5 years running at a 50 kton far
detector. We have computed them for $|\Delta m^{2}_{31}| = 2.4 \times 
10^{-3} \ \rm{eV}^2$, $\sin^2 2 \theta_{13} = 0.058$ and four values of
$\delta = 0, \frac{\pi}{2}, \pi, \frac{3\pi}{2}$. The remaining
oscillation parameters are $\Delta m^{2}_{21}= 8.0 \times 10^{-5} \
{\rm eV}^2$, $\sin^2 \theta_{12}= 0.31$, $\sin^2 2\theta_{23}= 1$ and
the matter parameter $A\equiv\sqrt{2} G_{F} n_e = 1.064 \times 10^{-4}
\ {\rm eV}^2 / {\rm GeV}$. The energy window is [1.8,2.8] ${\rm GeV}$.
The neutrino spectrum peaks at 2.3~${\rm GeV}$.}  
\label{tab:nova800}
\end{table}

\begin{table}[h]
\vspace{1cm}
\centering
\begin{tabular}{|c|c|c|c|c|c|}
\hline\hline
$\sin^{2} 2 \tetaot$ & $\delta$ & $\Delta m_{31}^2$ (eV$^2$)  &
$\nu_\mu$(unoscillated) &$\nu_e$ (signal) & $\nu_e$ (intrinsic
background)\\  
\hline\hline
0.058 & 0 & 0.0024& 13285 &342 &81\\
0.058 & $\pi$/2& 0.0024& 13285 &373 &81\\
0.058 & $\pi$ & 0.0024& 13285 &257 &81\\
0.058 & 3$\pi$/2 & 0.0024& 13285 & 226&81\\
\hline\hline 
0.058 & 0 &$ -0.0024$& 13285 &228 &81\\
0.058 & $\pi$/2 & $-0.0024$& 13285 &335 &81\\
0.058 & $\pi$ &$ -0.0024$& 13285 &301 &81\\
0.058 & 3$\pi$/2 & $-0.0024$& 13285 & 193&81\\
\hline\hline 
\end{tabular}
\caption{\it Same as Table~\ref{tab:nova800} for a baseline of 434 km
(10.23 km off-axis). The matter parameter $A\equiv\sqrt{2} G_{F} n_e =
8.93 \times 10^{-5} \ {\rm eV}^2 / {\rm GeV}$, and the energy window
is [0.8,1.8] ${\rm GeV}$. The neutrino spectrum peaks at 1.3~${\rm
  GeV}$.}
\label{tab:nova400}
\end{table}

\begin{table}[h]
\vspace{1cm}
\centering
\begin{tabular}{|c|c|c|c|c|c|}
\hline\hline
$\sin^{2} 2 \tetaot$ & $\delta$ & $\Delta m_{31}^2$ (eV$^2$)  &
$\nu_\mu$(unoscillated) &$\nu_e$ (signal) & $\nu_e$ (intrinsic
background)\\  
\hline\hline
0.058 & 0 & 0.0024& 8253 &163 &83\\
0.058 & $\pi$/2 & 0.0024& 8253 &177 &83\\
0.058 & $\pi$ & 0.0024& 8253 &122 &83\\
0.058 & 3$\pi$/2 & 0.0024& 8253 &108 &83\\
\hline\hline 
0.058 & 0 & $-0.0024$& 8253 &123 &83\\
0.058 & $\pi$/2 & $-0.0024$& 8253 &177 &83\\
0.058 & $\pi$ & $-0.0024$& 8253 &160 &83\\
0.058 & 3$\pi$/2 & $-0.0024$& 8253 &106 &83\\
\hline\hline 
\end{tabular}
\caption{\it Same as Table~\ref{tab:nova800} for a baseline of 200 km
(8.44 km off-axis). The matter parameter $A\equiv\sqrt{2} G_{F} n_e =
3.83 \times 10^{-5} \ {\rm eV}^2 / {\rm GeV}$, and the energy window
is [0.2,1.2] ${\rm GeV}$. The neutrino spectrum peaks at 0.7~${\rm
  GeV}$. }   
\label{tab:nova200}
\end{table}

\end{document}